\documentclass{aa}

\usepackage{natbib}
\usepackage{psfrag, color, epsfig, rotating, pstricks, pst-node}
\usepackage{txfonts}

\bibpunct{(}{)}{;}{a}{}{,}

\newlength{\hsizethird}
\font \bolditalics = cmmib10

\font \matrix = cmssdc10
\newcommand{\Vec}[1]{{\textfont1=\bolditalics \hbox{$#1$}}}
\newcommand{\Mat}[1]{{\textfont1=\matrix \hbox{$#1$}}}

\newcommand{\average}[1]{\langle #1 \rangle}

\newcommand{\vp}{\varphi}
\newcommand{\vt}{\vartheta}
\newcommand{\ve}{\varepsilon}
\newcommand{\enat}{{\cal E}}

\newcommand{\Cov}{\Mat{Cov}}

\newcommand{\dd}{{\mathrm d}}

\newcommand{\ontop}[2]{
  \renewcommand{\arraystretch}{0.2}
  \begin{array}{c}
  #1 \\ #2
  \end{array}
  \renewcommand{\arraystretch}{1.0}
}
\newcommand{\lsim}{\ontop{<}{\sim}}
\newcommand{\gsim}{\ontop{>}{\sim}}
\newcommand{\Map}{{M_{\rm ap}}}
\newcommand{\Mapsq}{M_{\rm ap}^2}

\newcommand{\del}{\partial}

\newcommand{\Omegam}{\Omega_{\rm m}}

\newcommand{\ii}{{\rm i}}
\newcommand{\ee}{{\rm e}}

\def\apjl{ApJ}%
\def\apjs{ApJS}%
\def\ao{Appl.~Opt.}%
\def\aap{A\&A}%
\begin{document}


\DeclareGraphicsExtensions{.eps, .ps}

\title{Cosmological parameters from combined second- and
       third-order aperture mass statistics of cosmic shear}

\titlerunning{Cosmological parameters from $\average{M_{\rm ap}^2}$
  and $\average{M_{\rm ap}^3}$}

\author{Martin Kilbinger\inst{1} \and Peter Schneider\inst{1}}

\institute{Institut f. Astrophysik
u. Extraterrestrische Forschung, Universit\"at Bonn, Auf dem H\"ugel 71,
D-53121 Bonn, Germany}

\offprints{Martin Kilbinger, \email{kilbinge@astro.uni-bonn.de}}

\date{Received / Accepted}

\abstract{
  We present predictions for cosmological parameter constraints from
  combined measurements of second- and third-order statistics of
  cosmic shear. We define the generalized third-order aperture mass
  statistics $\langle M_{\rm ap}^3 \rangle$ and show that it contains
  much more information about the bispectrum of the projected matter
  density than the skewness of the aperture mass. From theoretical
  models as well as from $\Lambda$CDM ray-tracing simulations, we
  calculate $\langle M_{\rm ap}^2 \rangle$ and $\langle M_{\rm ap}^3
  \rangle$ and their dependence on cosmological parameters. The
  covariances including shot noise and cosmic variance of $M_{\rm
    ap}^2$, $M_{\rm ap}^3$ and their cross-correlation are calculated
  using ray-tracing simulations. We perform an extensive Fisher matrix
  analysis, and for various combinations of cosmological parameters,
  we predict 1-$\sigma$-errors corresponding to measurements from a
  deep 29 square degree cosmic shear survey. Although the parameter
  degeneracies can not be lifted completely, the (linear) combination
  of second- and third-order aperture mass statistics reduces the
  errors significantly. The strong degeneracy between $\Omegam$ and
  $\sigma_8$, present for all second-order cosmic shear measures, is
  diminished substantially, whereas less improvement is found for the
  near-degenerate pair consisting of the shape parameter $\Gamma$ and
  the spectral index $n_{\rm s}$. Uncertainties in the source galaxy
  redshift $z_0$ increase the errors of all other parameters.
}

\maketitle

\section{Introduction}

In recent years, weak gravitational lensing by the large-scale matter
distribution in the Universe has become an important tool for
cosmology. Cosmic shear surveys have yielded constraints on
cosmological parameters without the need for modeling the relation
between luminous and dark matter (bias).
In particular, the power spectrum
normalization $\sigma_8$ has been obtained with less than 10 \%
uncertainty \citep{vWMH04}.

On the one hand, the observed sky area and thus the number of faint
background galaxies increased dramatically with the advent of
wide-field imaging cameras mounted onto large telescopes.  On the other
hand, measurement errors have decreased with further understanding of
systematics together with new image analysis methods. These two
advances were crucial in the evolution of cosmic shear towards a
high-precision cosmology tool.

Cosmic shear is sensitive to inhomogeneities in the projected matter
distribution out to redshifts of order unity, depending on the depth of
the survey. It probes scales where fluctuations have started to grow
non-linearly due to gravitational instabilities. These non-linearities
along with projection effects erase most of the primordial features
such as baryon wiggles in the power spectrum. Thus, cosmological
parameters cannot be determined uniquely using second-order
statistics alone; there exist substantial near-degeneracies, e.g.\ between
$\Omegam$ and $\sigma_8$.

Because these degeneracies manifest themselves in a different way for
shear statistics of different order, they can be lifted by combining
e.g.\ second- and third-order statistics. An example is the reduced
skewness of the convergence or projected surface mass density $\kappa$, which
has been shown to not, or only weakly, depend on $\sigma_8$ and thus to
be able to break the near-degeneracy with $\Omegam$
\citep{1997A&A...322....1B, 1999A&A...342...15V}.

Although the convergence cannot be observed directly,
\cite{1998MNRAS.296..873S} defined the so-called aperture mass
statistics $\Map$, which is a local convolution of $\kappa$ with a
compensated filter, and which can be measured directly from the
ellipticities of the background galaxies.


The first significant non-zero third-order cosmic shear signal was found by
\cite{BMvW02}, who measured an integral over the three-point
correlation function (3PCF) of shear in the VIRMOS-DESCART survey.
From the same data, aperture mass skewness was detected later by
\citet{2003ApJ...592..664P}, and an upper limit for $\Omega_\Lambda$
was derived.
A $\Map$ skewness detection at the $2\sigma$-level was obtained from the
CTIO survey by \cite{JBJ04} who also derived handy expressions for the
$\Map$ skewness in terms of the 3PCF. Both detections of $\langle M_{\rm
ap}^3 \rangle$ were obtained by integrating over the measured 3PCF.

In this paper, we demonstrate the improvement of cosmological
parameter determination from cosmic shear, using combined measurements
of the second- and generalized third-order aperture mass
statistics. This latter quantity was introduced by \cite{SKL04}
as the third-order correlator of $\Map$ for three
different aperture radii, $\average{M_{\rm ap}^3(\theta_1, \theta_2,
\theta_3)} \equiv \average{\Map(\theta_1) \Map(\theta_2)
\Map(\theta_3)}$. Unlike the skewness {$\average{M_{\rm ap}^3(\theta,
\theta, \theta)} \equiv \average{\Map(\theta) \Map(\theta)
\Map(\theta)}$, which depends on only one filter scale
$\theta$, the generalized third-order aperture mass statistics
contains information about the convergence bispectrum in principle
over the full Fourier-space.

The reasons of employing $\Map$ instead of the shear
correlation functions are multiple:
\begin{itemize}

\item $M_{\rm ap}$ is a scalar quantity, therefore odd powers of it
such as $M_{\rm ap}^3$ have non-trivial expectation
values. In contrast, no scalar can be formed from tri-linear
combinations of the eight components of the 3PCF of shear
\citep{tpcf1,2003ApJ...583L..49T,2003ApJ...584..559Z}.

\item From the aperture mass statistics, we get a measure of the
residual systematics by its ability to separate the E- from the B-mode
\citep{2002ApJ...568...20C,2002A&A...389..729S}. This is true for both
second- and third-order.

\item The integral relations between $\average{M_{\rm ap}^3}$ and the
bispectrum are much easier and numerically faster to evaluate
than for the 3PCF \citep{SKL04}. The reason is that $\average{M_{\rm
    ap}^3}$ is a local measure of the bispectrum, whereas the integral
kernel for the 3PCF is a highly oscillating function with infinite support.

\end{itemize}

This paper is organized as follows. In Sect.~\ref{sec:models}, we
describe the theoretical models that are employed for the power and
bispectrum.  We give the definition of the second- and generalized
third-order aperture mass statistics and their relation to the power
and bispectrum in Sect.~\ref{sec:map}, followed by a short description
of the ray-tracing simulations. Section \ref{sec:cov} addresses the
calculation of the covariance matrices of ${M_{\rm ap}^2}$, ${M_{\rm
ap}^3}$ and their cross-correlation.  Finally, in
Sect.~\ref{sec:constraints} we present our results on cosmological
parameter constraints from a Fisher matrix analysis.

\section{Models of the power- and bispectrum}
\label{sec:models}

Statistical weak gravitational lensing on large scales
probes the projected density field of the matter in the Universe, also
called convergence $\kappa$.
All second-order statistics of the convergence can be expressed as
functions of the two-point correlation function (2PCF) of $\kappa$ or its
Fourier transform, the power spectrum $P_\kappa$.
Analogously, third-order statistics are related to the 3PCF of
$\kappa$; its Fourier transform is the bispectrum $B_\kappa$.

The 3-D dark matter power spectrum has been
extensively modeled using numerical simulations. Halo model approaches
as well as fitting formulae give very accurate descriptions of the
quasi-linear and highly non-linear regime on intermediate and small
scales \citep{PD96,
2003MNRAS.341.1311S,2002PhR...372....1C}. Throughout this work, we
employ the fitting formula of \cite{PD96}, which was also used by
\cite{2001MNRAS.325.1312S} for their modeling of the bispectrum.

On the other hand, the bispectrum of the cosmological dark matter
distribution is less securely known.  It is well established that the
primordial density fluctuations were Gaussian
\citep[e.g.][]{Spergel03}. In the limit of linear perturbations, they
remain Gaussian -- thus the power spectrum alone contains all
information about the large-scale structure. However, gravitational
clustering is a non-linear process and, in particular at small scales,
the mass distribution evolves to become highly non-Gaussian.


The bispectrum $B_\kappa$ of the convergence is defined by the
following equation:
\begin{eqnarray}
\lefteqn{\left\langle\hat \kappa(\vec \ell_1) \hat \kappa(\vec \ell_2) \hat \kappa(\vec
\ell_3)\right\rangle  = (2 \pi)^2 \delta_{\rm D}(\vec \ell_1 + \vec \ell_2 + \vec \ell_3)}
\nonumber \\
& & \hspace{3em} \times \left[ B_\kappa(\vec \ell_1, \vec \ell_2) +
B_\kappa(\vec
\ell_2, \vec \ell_3) + B_\kappa(\vec \ell_3, \vec \ell_1)
\right],
\end{eqnarray}
where $\hat \kappa$ is the Fourier transform of $\kappa$ and
$\delta_{\rm D}$ is Dirac's delta function.

We assume the field $\kappa$ to be statistically isotropic, thus its
bispectrum only depends on the moduli $\ell_1, \ell_2$ of the wave
vectors and their enclosed angle $\vp$, $B_\kappa(\vec \ell_1, \vec
\ell_2) \equiv b_\kappa(\ell_1, \ell_2, \vp)$. Because of parity
symmetry, $b_\kappa$ is an even function of $\vp$.

In this work, we employ hyper-extended perturbation theory
\citep[HEPT,][]{2001MNRAS.325.1312S} for a $\Lambda$CDM Universe as a
model for the bispectrum. The HEPT fitting formula fits the $N$-body
simulations with an accuracy of $\lsim$ 15 percent,
which is sufficient for our purpose. In HEPT, we can write
\begin{equation}
b_\kappa(\ell_1, \ell_2, \vp) = \sum_{m=0}^2 F_2^{(m)}(\ell_1, \ell_2)
\cos^m (\vp) \, \bar b_\kappa^{(m)}(\ell_1, \ell_2),
\label{bkappa}
\end{equation}
with $F_2^{(0)} = 10/7$, $F_2^{(1)} = \ell_1/\ell_2 + \ell_2/\ell_1$,
$F_2^{(2)} = 4/7$. The functions $\bar b_\kappa^{(m)}$ are projections
of the 3-D bispectrum of density fluctuations $\delta$ which in the
quasi-linear regime are given in terms of the power spectrum
$P_\delta$. The projection is calculated using Limber's equation, and yields
\begin{eqnarray}
\lefteqn{\bar b^{(m)}_\kappa(\ell_1, \ell_2) = \int\limits_0^{w_{\rm
      hor}} \frac{\dd w}{f_K(w)} G^3(w) f^{(m)}(w, \ell_1) f^{(m)}(w,
  \ell_2)} \nonumber \\ & & \hspace{3em} \times
P_\delta\left(\frac{\ell_1}{f_K(w)}\right)
P_\delta\left(\frac{\ell_2}{f_K(w)}\right).
\label{bkappabar}
\end{eqnarray}
Here, $f_K(w)$ is the comoving angular distance ($f_K \equiv \rm{id}$ for
a flat Universe) and $w$ is the comoving distance.
The lens efficiency function $G$ is
\begin{equation}
G(w) = \frac 3 2 \left( \frac {H_0} c \right)^2 \frac \Omegam {a(w)}
\int\limits_w^{w_{\rm hor}} \dd w^\prime p(w^\prime) \frac
{f_K(w^\prime - w)}{f_K(w^\prime)},
\label{G}
\end{equation}
where $p$ denotes the probability distribution of the comoving number
density of source galaxies. For the ray-tracing simulations we
will assume that all source galaxies are at redshift $z_0 \approx 1$.

In quasi-linear perturbation theory (PT), $f^{(0)} = f^{(1)} = f^{(2)}
= 1.$ In HEPT however, we have to insert for $f^{(m)}, \, m=0,1,2$ the
fitting functions $a$, $b$ and $c$ respectively, as given in eqs.\ (10
- 12) of \cite{2001MNRAS.325.1312S}. These coefficients depend on the
wave vector $\ell$ measured in units of some non-linear scale
$\ell_{\rm NL}(w)$, the local spectral index of the linear power
spectrum $n(\ell)$ and weakly on the power spectrum normalization
$\sigma_8$ and the linear growth factor. The HEPT fitting functions
$a$, $b$ and $c$ parametrize a non-linear generalization of PT and
were obtained by \cite{2001MNRAS.325.1312S} using $N$-body
simulations. In the large-scale limit, these functions approach unity
to recover the PT results. For very small scales, $a$ is constant, $b$
and $c$ vanish, so that the bispectrum (\ref{bkappa}) becomes
independent of $\vp$ and thus the reduced bispectrum, which is
basically the ratio of $b_\kappa$ and the square of the power
spectrum, becomes independent of the triangle configuration and takes
the value of the hierarchical amplitude of stable clustering.

For the sake of completeness, we also give the power spectrum of the
convergence,
\begin{equation}
  P_\kappa(\ell) = \int \dd w \, G^2(w) P_\delta\left({\ell} \over
  {f_K(w)}\right). 
\end{equation}

\medskip

\section{Second- and third-order aperture mass}
\label{sec:map}

\subsection{Definition}

The aperture mass, introduced by \cite{KSFW94} and
\cite{S96}, is defined as the integral over the filtered surface mass
density $\kappa$ in an aperture, centered at some point $\vec \vt$.
Alternatively, it can be expressed in terms of the tangential shear
$\gamma_{\rm t}(\vec \vt^\prime) = - \Re [ \gamma(\vec \vt^\prime)
\exp(-2\ii\vp) ]$, where $\vp$ is the polar angle of the vector $\vec
\vt^\prime - \vec \vt$, such that the tangential component of the shear is
understood with respect to the aperture center $\vec \vt$. With a filter
function $U_\theta$, the definition reads
\begin{eqnarray}
\Map(\theta, \vec \vt) & = & \int \dd^2 \vt^\prime \, U_\theta(| \vec \vt -
\vec \vt^\prime|) \, \kappa(\vec \vt^\prime) \nonumber \\
        & = & \int \dd^2 \vt^\prime \, Q_\theta(| \vec \vt -
        \vec \vt^\prime|) \, \gamma_{\rm t}(\vec \vt^\prime),
\label{map_defx}
\end{eqnarray}
the second equality holds if $U_\theta$ is a compensated filter
function, i.e.\ $\int \dd \vt \, \vt \, U_\theta(\vt) = 0$, and
\begin{equation}
Q_\theta(\vt) = \frac 2 {\vt^2} \int\limits_0^\vt \dd \vt^\prime \, \vt^\prime \,
U_\theta(\vt^\prime) - U_\theta(\vt).
\end{equation}
It is the second equality in (\ref{map_defx}) which makes the aperture mass
statistics so useful, because it can be estimated by averaging over
the (weighted) tangential ellipticities in an aperture.
The integrals (\ref{map_defx}) can be written as convolution,
\begin{equation}
\Map(\theta, \vec \vt) = \left( U_\theta * \kappa
\right)(\vec \vt) = \Re \left( Q^\prime_\theta * \gamma \right)(\vec \vt),
\label{map_conv}
\end{equation}
where we defined the modified filter function
\begin{equation}
Q^\prime_\theta(\vt) = - Q_\theta(\vt) \ee^{-2\ii \arctan(\theta_2/\theta_1)}.
\end{equation}

The first moment of (\ref{map_conv}) vanishes, because of the
compensated nature of the filter $U_\theta$.

The second moment or dispersion of (\ref{map_conv})
\citep{1998MNRAS.296..873S} has been measured with great success in
numerous cosmic shear surveys
\citep[e.g.][]{vWMH04,J03,2002ApJ...577..595H,2003ApJ...597...98H}. Because it
separates the E- from the B-mode, it is an extremely useful tool to assess
measurement errors and systematics. Moreover, $\langle M_{\rm
ap}^2\rangle$ is a local measure of the power spectrum and therefore
very sensitive to cosmological parameters.

The next-higher order quantity is the third moment or skewness
of (\ref{map_conv}) \citep{1998MNRAS.296..873S,JBJ04}.  However, a
logical step is to generalize this statistics and allow for
correlations on different filter scales $\theta_1, \theta_2$ and
$\theta_3$ \citep{SKL04}. We denote this new quantity with
$\average{M_{\rm ap}^3(\theta_1, \theta_2, \theta_3)} \equiv
\average{\Map(\theta_1) \Map(\theta_2) \Map(\theta_3)}$, in contrast
to the case of three equal filter scales, $\langle M_{\rm ap,
d}^3(\theta)\rangle \equiv \average{M_{\rm ap}^3(\theta, \theta,
\theta)}$.

We expect the generalized aperture mass to carry much more information
than the `diagonal' case $\langle M_{\rm ap, d}^3 \rangle$. The latter
basically samples the bispectrum for equilateral triangles only,
whereas $\average{M_{\rm ap}^3}$ probes the bispectrum essentially over
the full $\ell$-space, as was shown
in \cite{SKL04}, see also Fig.~\ref{fig:int-bk}.

Throughout this paper, we employ the filter functions given by
\cite{2002ApJ...568...20C},
\begin{equation}
U_\theta(\vt) = \frac 1 {2\pi\theta^2} \left( 1 - \frac{\vt^2} {2\theta^2}
\right) \ee^{- \frac{\vt^2} {2\theta^2}} ; \; Q_\theta(\vt) =
\frac{\vt^2}{4\pi\theta^4} \ee^{- \frac{\vt^2} {2\theta^2}}.
\end{equation}

The disadvantage of these functions is their infinite
support. Although decreasing exponentially, they are significantly
non-zero up to about three times the aperture radius
$\theta$. However, the usage of filter functions with finite support,
e.g.\ the polynomial filters from \cite{1998MNRAS.296..873S}, would
involve much more cumbersome expressions for $\langle M_{\rm
ap}^3\rangle$ as a function of the 3PCF, which makes the aperture mass
statistics very unhandy when it has to be inferred from real
data.

\subsection{Theoretical calculations of the aperture mass statistics}

\begin{figure*}
  \resizebox{\hsize}{!}{
    \includegraphics{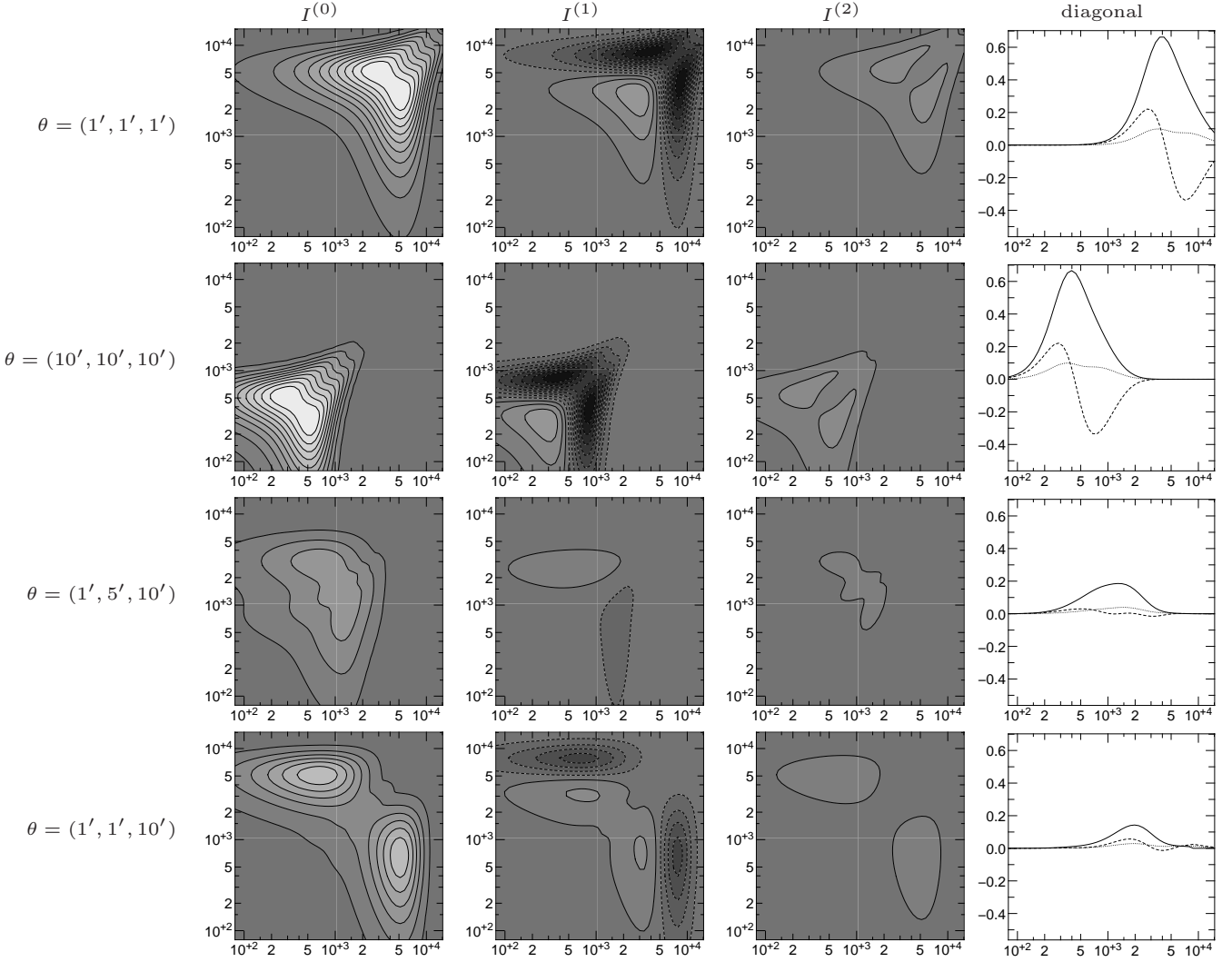}}
  \caption{The filter functions $I^{(m)}$ (\ref{Im}) for the generalized
    third-order aperture mass statistics as a function of the bispectrum
    (\ref{Map3-bisp}). Contours of the $I^{(m)}$ for $m=0,1,2$ from left
    to right are plotted for different values of $\theta_i$ as a function
    of $\ell_1$ and $\ell_2$ (in units of inverse radians).  The dashed
    contours indicate negative values.  The right-most panels show the
    profile along the diagonal of $I^{(m)}$, where the solid, dashed and
    dotted lines correspond to $m=0,1$ and 2 respectively.}
  \label{fig:int-bk}
\end{figure*}

The second- and third-order aperture mass statistics can be calculated
as integrals over the power spectrum and the bispectrum of the
convergence $\kappa$, respectively.
For second order, we have 
\begin{equation}
\left\langle{M_{\rm ap}^2(\theta)}\right\rangle = \int \frac{\dd \ell \, \ell}{2\pi}
P_\kappa(\ell) \hat U^2(\theta \ell),
\label{map2-power}
\end{equation}
where $\hat U(\theta \ell) = {\cal F}[U_\theta](\ell) = (\theta
\ell)^2/2 \cdot \exp[-(\theta \ell)^2/2]$ is the Fourier transform of
the filter function $U_\theta$.  The generalized third-order aperture
mass statistics can be written as
\begin{eqnarray}
\lefteqn{\left\langle{M_{\rm ap}^3(\theta_1, \theta_2, \theta_3)}\right\rangle \equiv
\average{M_{\rm ap}(\theta_1)M_{\rm ap}(\theta_2)M_{\rm ap}(\theta_3)}}
\nonumber \\
& = & \int \frac{\dd^2\ell_1}{(2\pi)^2}
\int \frac{\dd^2\ell_2}{(2\pi)^2}\,B_\kappa(\vec\ell_1,\vec\ell_2)
\nonumber \\
 & & \times \sum\limits_{(i,j,k) \in S_3} \hat
U(\theta_i|\vec\ell_1|)\,\hat U(\theta_j|\vec\ell_2|)\,
\hat U(\theta_k |\vec\ell_1+\vec\ell_2|),
\label{map3-bi}
%
%
\end{eqnarray}
where $S_3$ is the symmetric permutation group of $(123)$, thus the
summation is performed over even permutations of $i,j,k$ \citep{SKL04}.

Both integrals (\ref{map2-power}) and (\ref{map3-bi}) are easily
calculated numerically due to the exponential cut-off of $\hat U$ for
large $\ell$. Eq.\ (\ref{map3-bi}) can be simplified further if the
bispectrum can be factorized as in (\ref{bkappa}). Then, terms of the
form
%
\begin{eqnarray}
\lefteqn{
K^{(m)}(\theta \ell_1, \theta \ell_2) =
}  \nonumber \\
& & 
\hspace{2em}
\int\limits_0^{2\pi} \dd \vp
\cos^m (\vp) \,
\hat U\left(\theta \sqrt{\ell_1^2 + \ell_2^2 + 2\ell_1 \ell_2
\cos\vp }\right),
\end{eqnarray}
%
for $m=0,1,2$ can be separated and carried out analytically,
\begin{eqnarray}
K^{(0)}(t_1, t_2) & = & \pi \ee^{-\frac 1 2 (t_1^2 +
        t_2^2)} \left[ (t_1^2 + t_2^2) {\rm I}_0(t_1 t_2
        ) 
  - 2 t_1 t_2 {\rm I}_1(t_1 t_2) \right], \nonumber\\
K^{(1)}(t_1, t_2) & = & \pi \ee^{-\frac 1 2 (t_1^2 +
        t_2^2)} \left[ 2 t_1 t_2 {\rm I}_0(t_1 t_2) \right. \nonumber
        \\
        & & \left. - (2 + t_1^2 + t_2^2) {\rm I}_1(t_1 t_2) \right]
\label{Km} \\
K^{(2)}(t_1, t_2) & = & \pi\ee^{-\frac 1 2(t_1^2 +
        t_2^2)} \bigg[ (2 + t_1^2 + t_2^2) {\rm I}_0(t_1 t_2)
        \nonumber \\
        & & \left. - \left( \frac{t_1^2 + t_2^2 + 4}{t_1 t_2} + 2 t_1 t_2
        \right) {\rm I}_1(t_1 t_2) \right], \nonumber
\end{eqnarray}
where ${\rm I}_n$ is the modified Bessel function of order $n$. We get
for (\ref{map3-bi})
\begin{eqnarray}
  \lefteqn{\left\langle{M_{\rm ap}^3(\theta_1, \theta_2, \theta_3)}\right\rangle =
    (2\pi)^{-3} \int \dd \ell_1 \, \ell_1 \int \dd \ell_2 \,
    \ell_2} \nonumber \\ 
  & & \hspace{3em} \times \sum_{m=0}^2
  I^{(m)}(\theta_1, \theta_2, \theta_3; \ell_1, \ell_2) \, \bar
  b_\kappa^{(m)}(\ell_1, \ell_2),
\label{Map3-bisp}
\end{eqnarray}
with
\begin{eqnarray}
  \lefteqn{I^{(m)}(\theta_1, \theta_2, \theta_3; \ell_1, \ell_2) =
    F_2^{(m)}(\ell_1, \ell_2)} \nonumber \\ 
  & & \hspace{3em} \times \sum\limits_{(i,j,k)
    \in S_3} \hat U(\theta_i \ell_1) \, \hat U(\theta_j \ell_2)
    K^{(m)}(\theta_k \ell_1, \theta_k \ell_2).
\label{Im}
\end{eqnarray}
One sees in Fig.~\ref{fig:int-bk} that the functions $I^{(m)}$ as
defined in the previous equation are relatively well localized which
makes $\average{M_{\rm ap}^3}$ a local measure of the bispectrum.
Further, for equal filter scales, only the region around the diagonal
of the $\bar b_\kappa^{(m)}$ is probed, corresponding to equilateral
triangles in Fourier space.  When different filter scales are taken
into account, other parts further away from the diagonal of $\bar
b_\kappa^{(m)}$ can contribute to the integral. Although in this
latter case the amplitude of $I^{(m)}$ is lower, the generalized
aperture mass, probing the bispectrum for general triangles in Fourier
space, contains much more information about the bispectrum and
cosmology than the `diagonal' one. This approach to sample the
bispectrum on a large region of Fourier space is similar to a previous
study \citep{2004MNRAS.348..897T}, who have used all triangle
configurations of the convergence bispectrum in order to predict tight
constraints on cosmological parameters from cosmic shear. In contrast
to that work, we use moments of the aperture mass statistics (which
are direct weak lensing observables) as real-space probes of the
convergence power spectrum and bispectrum.

\subsection{Ray-tracing simulations}
\label{sec:ray}

We use 36 $\Lambda$CDM ray-tracing simulations, kindly provided by
T. Hamana \citep[for more details see][]{2003A&A...403..817M} in order
to calculate the second- and third-order aperture mass statistics and
their covariances (Sect.~\ref{sec:cov}) . Each field consists of
$1024^2$ data points in $\kappa$ and $\gamma$, the pixel size is
$0.2^\prime$. We assume our galaxies to be given on a regular grid --
every pixel corresponds to a galaxy, thus our source galaxy density is
25 per square arc minute. We note here that the Poisson noise is much
smaller than the shape noise of the ellipticities, and that apertures
with radii smaller than one arc minute are discarded due to
discreteness effects in the ray-tracing and in the underlying $N$-body
simulations.

All source galaxies are located at a redshift of about unity. See
Table \ref{tab:param} for the fiducial values of the parameters.

\begin{table}[!b]
  \caption{Fiducial values of the cosmological parameters that are
    used for the theoretical model to match the ray-tracing
    simulations.  If the shape parameter $\Gamma$ is interpreted as
    Sugiyama's $\Gamma$ \citep{1995ApJS..100..281S}, our fiducial
    model corresponds to $\Omega_{\rm b} = 0.04$ and $h = 0.7$.
  }
  \begin{center}
  \begin{tabular}{cccccc}\hline\hline
    $\Omegam$ & $\Omega_\Lambda$ & $\Gamma$ & $\sigma_8$ & $n_{\rm
      s}$ & $z_0$ \\ \hline
    0.3 & 0.7 & 0.1723 & 0.9 & 1.0 & 0.9772 \\ \hline\hline
  \end{tabular}
  \end{center}
  \label{tab:param}
\end{table}

\begin{figure}[t]
  \begin{center}
    \resizebox{\hsize}{!}{
      \includegraphics{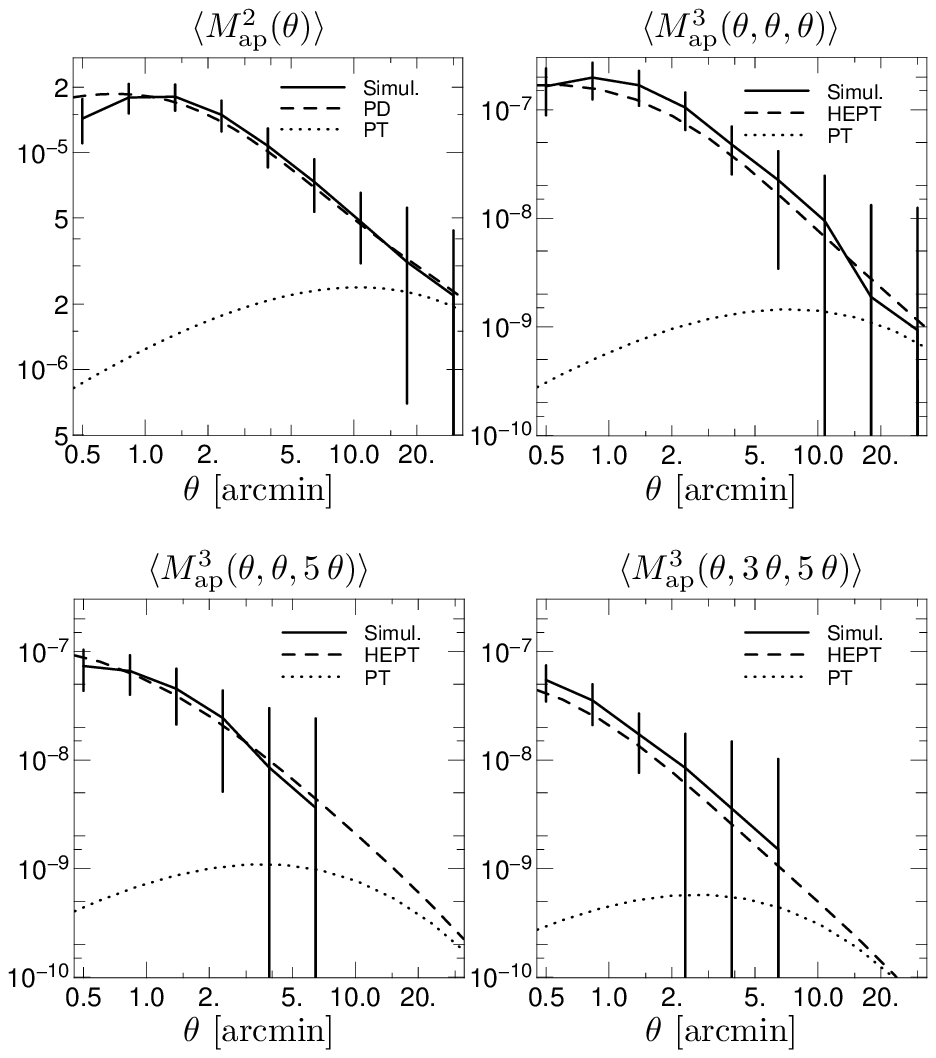}
    }
  \end{center}
  \caption{$\average{M_{\rm ap}^2}$ and $\langle M_{\rm ap}^3 \rangle$
    from the 36 $\Lambda$CDM simulations (solid lines) as compared to
    theoretical predictions (dashed and dotted lines). The error bars are
    the rms values from the 36 fields. $\langle M_{\rm
      ap}^3 \rangle$ is calculated from the simulations for aperture
    radii smaller than one sixth of the field size. PD = \cite{PD96},
    HEPT = \cite{2001MNRAS.325.1312S}, PT = (quasi-)linear perturbation
    theory.}
  \label{fig:map2_comp}
\end{figure}

Because the field $\kappa$ is given on a regular grid, moments of the
aperture mass statistics (\ref{map_conv}) can be calculated very
quickly using FFT, with the ensemble averages replaced by the average
over all aperture centers $\vec \vt$. However, since for discrete
Fourier transforms, periodic boundary conditions are assumed, which is
not the case for the ray-tracing simulations, points near the borders
have to be excluded from the averaging. This leads to smaller
effective area and therefore to an overestimation of the covariance of
the $M_{\rm ap}$-statistics, which increases with the aperture radius.
In order to avoid this, one could calculate $\average{M_{\rm ap}^2}$
and $\langle M_{\rm ap}^3 \rangle$ from the shear correlation
functions, which takes into account the complete area. This approach
is not chosen here because of the time-consuming calculation of the
3PCF. The correction scheme we apply to the covariance matrices is
described in Sect.~\ref{sec:cov-not}.

Figs.~\ref{fig:map2_comp} and \ref{fig:map3_cont_ray+num} show
$\average{M_{\rm ap}^2}$ and $\average{M_{\rm ap}^3}$ from the
$\Lambda$CDM simulations and the theoretical predictions based on
HEPT. The non-linear fitting formulae reproduce reasonably well the
results from the simulations for angular scales above $\sim$ 1 arc
minute. The largest aperture which can be put onto the field without
being too close to the border is for $\theta_{\rm max} =
a/6 = 34^\prime$, where $a = 204.8^\prime$ is the field size.

\begin{figure}[t]
  \resizebox{\hsize}{!}{
    \includegraphics{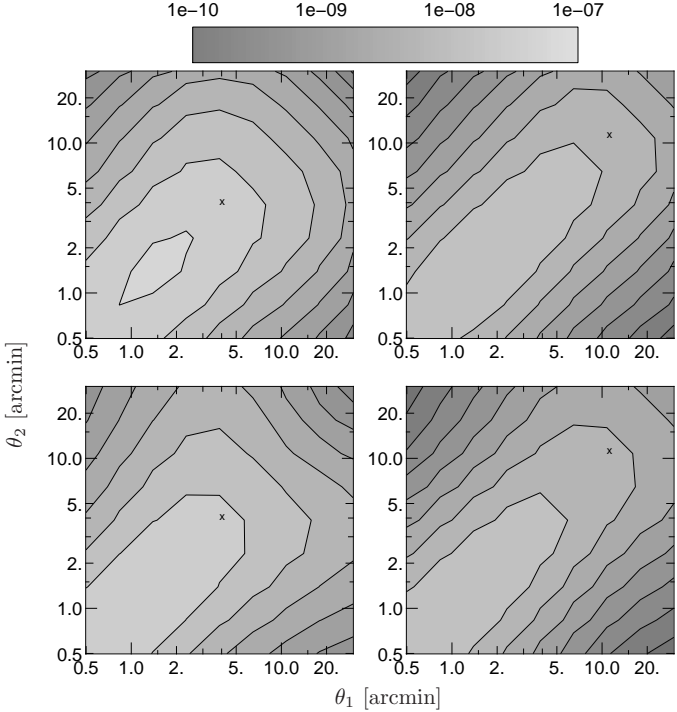}}
  \caption{Contours of $\average{M_{\rm ap}^3(\theta_1, \theta_2,
      \theta_3)}$ from simulations (upper row) and from the HEPT model
    (lower row). In each panel $\theta_3$ is fixed to the value
    indicated by the cross. These are 3.87 arc minutes (left column) and
    10.77 arc minutes (right column).}
  \label{fig:map3_cont_ray+num}
\end{figure}

For comparison, we calculate $\average{M_{\rm ap}^3}$ by integrating
over the 3PCF, using eqs.\ (62) and (71) from \cite{SKL04}. Although
we use the fast tree-code algorithm of \cite{JBJ04} to calculate the
3PCF, it is still very time-taking since a fine binning of the 3PCF is
needed (see below). Our results are shown in Fig.~\ref{fig:3pcf} and
represent the average over three of the ray-tracing fields.
$\average{M_{\rm ap}^3}$ as calculated via apertures cannot be
determined for large radii because of the border effects, as mentioned
above. Since $\average{M_{\rm ap}^3}$ obtained via integrating over
the 3PCF is based on the simulated shear field, we use the $\gamma$
fields instead of the $\kappa$ fields in order to calculate
$\average{M_{\rm ap}^3}$ via the FFT aperture method, using the second
equality in (\ref{map_conv}). With $M_\perp(\theta) = \Im \left(
  Q^\prime_\theta * \gamma \right)$, we also determine the statistics
$\langle M_{\rm ap}^2 M_\perp^{} \rangle$, $\langle M_{\rm ap}
M_\perp^2 \rangle$ and $\langle M_\perp^3 \rangle$ as indicators of a
B-mode. $\langle M_{\rm ap} M_\perp^2 \rangle$ is expected to vanish
if the ray-tracing simulations are B-mode-free. The two quantities
with odd power in $M_\perp$ can only be non-zero for a convergence
field which is not parity-invariant \citep{2003A&A...408..829S}. We
found all three statistics to be three and more orders of magnitude
below the pure E-mode, confirming that the ray-tracing simulations
contain virtually no B-mode and are parity-invariant. However, when
inferred from the 3PCF, $\langle M_{\rm ap} M_\perp^2 \rangle$ is at a
couple of percent of the E-mode. This is most probable due to the
binning of the 3PCF --- the B-mode gets smaller when we refine the
binning. In our calculations, we use a logarithmic bin width of
$b=0.075$.  As can be seen in Fig.~\ref{fig:3pcf}, there is good
agreement between the two methods, except for very small angular
scales (where the B-mode is of the order 10\%) and large aperture
radii (where a significant fraction of the field near the border can
not be taken into account with the aperture method).

\begin{figure}
  \begin{center}
    \resizebox{0.7\hsize}{!}{
      \includegraphics{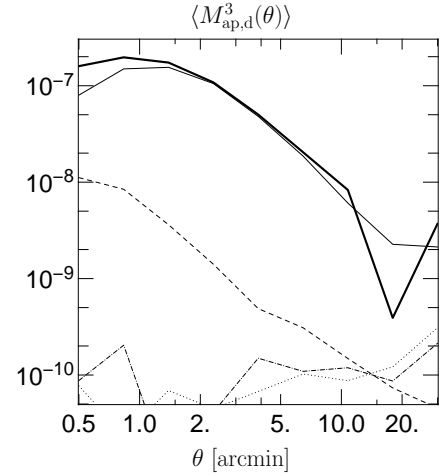}
    }
  \end{center}
  \caption{$\average{M_{\rm ap,d}^3}(\theta)$ from apertures (bold line), and
    from the integration over the 3PCF (thin line). Also plotted are
    the B-mode signals from the integration method $\langle M_{\rm ap}
    M_\perp^2 \rangle$ (dashed), $\langle M_{\rm ap}^2 M_\perp \rangle$
    (dotted) and $\langle M_\perp^3 \rangle$ (dash-dotted). The curves
    represent the mean from three of the ray-tracing fields.}
  \label{fig:3pcf}
\end{figure}

Integrating over the 3PCF is the preferred method in the case of real
data, since the determination of correlation functions is not affected
by unusable regions which makes placing apertures onto the observed
field very ineffective. However, the calculation of the 3PCF is very
time-consuming even using the fast tree-code algorithm. Moreover, a
relatively fine binning of the 3PCF is needed in order not to
introduce a B-mode from the integration of the 3PCF, and the
computation time goes as $b^{-3.3}$ \citep{JBJ04} where $b$ is the
logarithmic bin width. With $b=0.075$, the integration method takes
about a factor 500 longer than the aperture method using FFT.


\subsection{Dependence on cosmological parameters}

The goal of this paper is to study the ability of weak lensing
measurements of the aperture mass statistics to constrain cosmological
parameters. It is therefore instructive to show the dependence of the
aperture mass on various cosmological parameters, and to compare its
second- and third-order moments. The more different the dependencies
are for the second- and third-order statistics, the better will be the
improvement on the parameter constraints when combining both.

\begin{figure}[!t]
  \resizebox{\hsize}{!}{
    \includegraphics{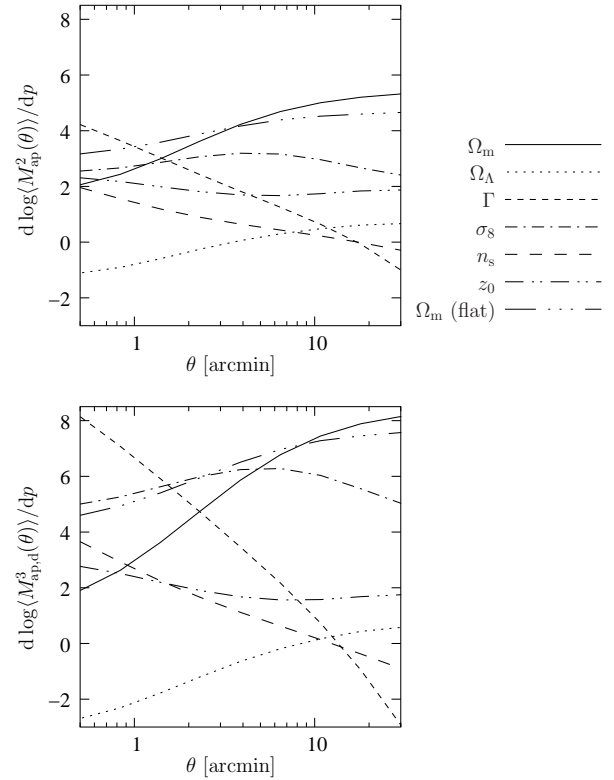}
  }
  \caption{Logarithmic derivatives of $\langle M_{\rm ap}^2\rangle$ and
    $\langle M_{\rm ap,d}^3\rangle$ with respect to some cosmological
    parameters as indicated in the figure legend.}
  \label{fig:dMap_dp}
\end{figure}

\begin{figure}[!ht]
  \resizebox{\hsize}{!}{
    \includegraphics{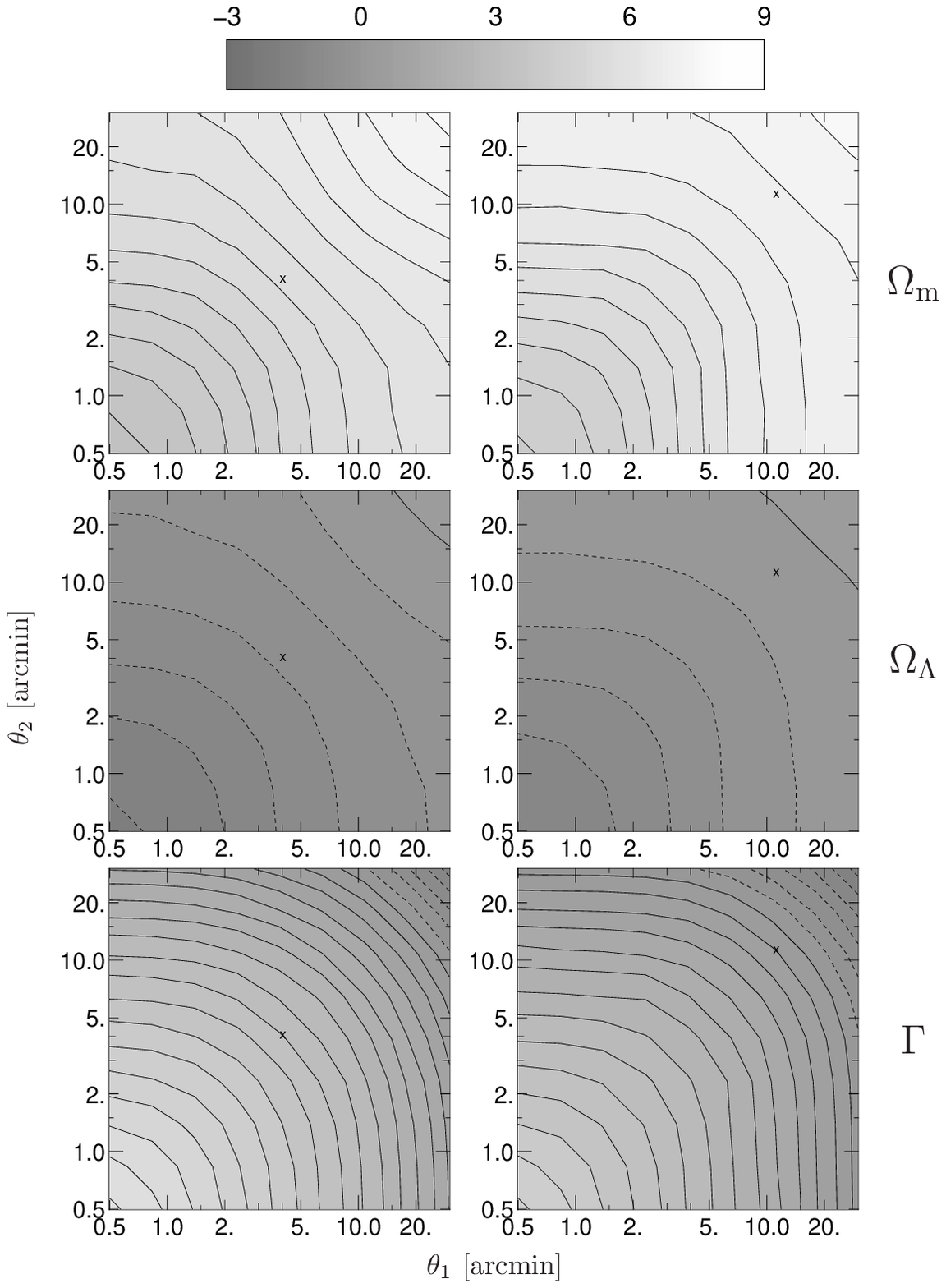}
  }
  \caption{Contours of $\dd \log \average{M_{\rm ap}^3(\theta_1,
      \theta_2, \theta_3)} / \dd p$ from the HEPT model, with
    $p=\Omegam, \Omega_\Lambda, \Gamma$ from
    top to bottom.
    In each panel $\theta_3$ is fixed to the value indicated by the
    cross. These are 3.87 (left column) and 10.77 arcminutes (right column).}
  \label{fig:dmap3_dp1_cont}
\end{figure}

\begin{figure}[!ht]
\resizebox{\hsize}{!}{
  \includegraphics{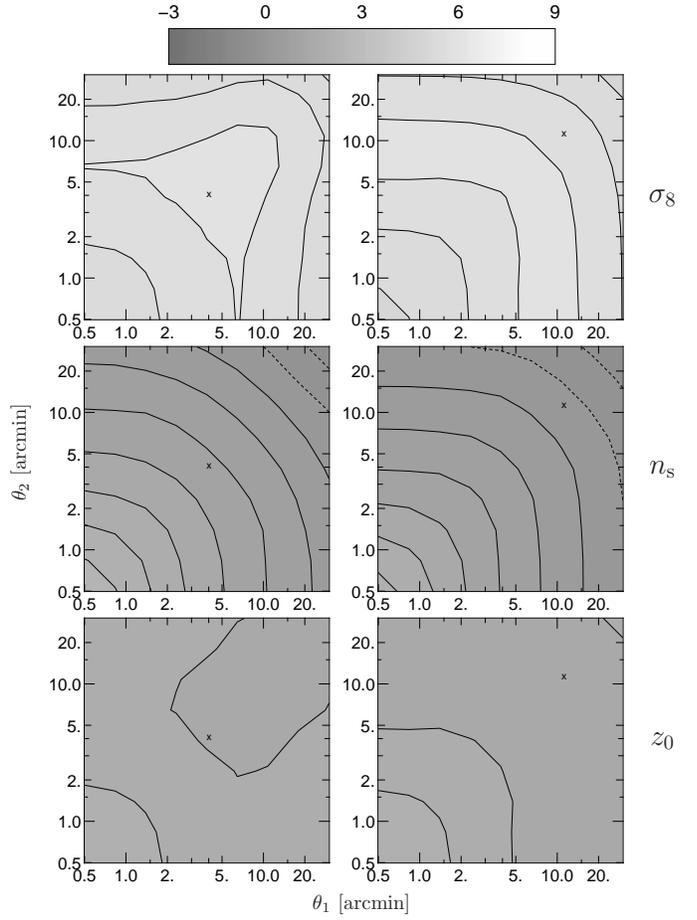}
}
\caption{Contours of $\dd \log \average{M_{\rm ap}^3(\theta_1,
    \theta_2, \theta_3)} / \dd p$ from the HEPT model, with
    $p=\sigma_8, n_{\rm s}, z_0$ from
    top to bottom.
    In each panel $\theta_3$ is fixed to the value indicated by the
    cross. These are 3.87 (left column) and 10.77 arcminutes (right column).}
\label{fig:dmap3_dp2_cont}
\end{figure}

In Figs.~\ref{fig:dMap_dp} - \ref{fig:dmap3_dp2_cont}, the
logarithmic derivatives of the aperture mass statistics with respect
to cosmological parameters used here are shown.
In all cases, the curves shown in Fig.\ \ref{fig:dMap_dp} are quite
featureless, their similarity is due to the near-degeneracies between
the parameters. For example, we find that the ratios ($\del \langle M_{\rm ap}^2
\rangle / \del n_{\rm s}) /( \del \langle M_{\rm ap}^2 \rangle / \del
\Gamma)$ $\approx$ $(\del \langle M_{\rm ap, d}^3 \rangle / \del
n_{\rm s}) /( \del \langle M_{\rm ap, d}^3 \rangle / \del \Gamma)$ are
roughly equal and constant as a function of the aperture radius
$\theta$. Therefore, we expect these two parameters to have the same
near-degeneracy for both statistics.

The ratio of derivatives with respect to $\Omegam$ and $\sigma_8$ are
slowly increasing functions of $\theta$, with significant differences
between $\langle M_{\rm ap}^2 \rangle$ and $\langle M_{\rm ap,d}^3
\rangle$.  From that we can infer that the reduced skewness $s_3 =
\langle M_{\rm ap,d}^3 \rangle / \langle M_{\rm ap}^2 \rangle^2$
breaks the $\Omegam$ - $\sigma_8$ degeneracy of second-order cosmic
shear statistics.  From Fig.~\ref{fig:dMap_dp} one sees that $\del
\log \langle M_{\rm ap,d}^3 \rangle / \del \sigma_8 \approx 2 \del
\log \langle M_{\rm ap}^2 \rangle / \del \sigma_8$, so $\del s_3 /
\del \sigma_8 \approx 0$ --- $s_3$ is indeed nearly independent of
$\sigma_8$, as predicted from quasi-linear perturbation theory
\citep{1997A&A...322....1B,1998MNRAS.296..873S}.

\section{Covariance matrices of aperture mass statistics}
\label{sec:cov}

\subsection{Definition and Notation}
\label{sec:cov-not}

Let $M_i$ be an estimator of some statistics, e.g.\ of the
second-order aperture mass $\langle M_{\rm ap}^2(\theta_i) \rangle$ for
some aperture radius $\theta_i$.  The covariance matrix of this
estimator is defined as
\begin{equation}
  {\rm Cov}(M)_{ij} = \left\langle{M_i M_j}\right\rangle -
  \big\langle{M_i}\big\rangle \left\langle{M_j}\right\rangle.
  \label{cov-def}
\end{equation}
In case of the generalized third-order aperture mass $\average{M_{\rm
ap}^3(\theta_i, \theta_j, \theta_k)}$, the covariance depends on six
scalar quantities, namely the $2 \times 3$ filter scales involved. In
order to obtain a two-dimensional matrix, we relabel all
non-degenerate combinations of filter triplets ($\average{M_{\rm
ap}^3}$ is invariant under permutations of its arguments) with a
single index.  The resulting $\langle M_{\rm ap}^3(\theta_1, \theta_2,
\theta_3) \rangle$-vector is organized such that $(\theta_1, \theta_2,
\theta_3)$ is in lexical order, we further demand that $\theta_1 \le
\theta_2 \le \theta_3$.  Note that the labeling order does not play a
role in the later analysis. For a number of $N$ distinct filter
scales, there are ${N+2 \choose 3} = N(N+1)(N+2)/6$ different
combinations.

We define the two covariance matrices $\Cov(M_{\rm ap}^2)$ and
$\Cov(M_{\rm ap}^3)$ for the second- and generalized third-order
aperture mass statistics, respectively. Further, for the skewness of
$M_{\rm ap}$, which is a function of only one
filter scale, $M_{\rm ap, d}^3(\theta) \equiv M_{\rm ap}^3(\theta,
\theta, \theta)$, we define the covariance matrix $\Cov(M_{\rm ap,
d}^3)$.

The averaging in eq.\ (\ref{cov-def}) is performed over the different
simulations. Because of the small number of realizations, we split up
each of the 36 fields into 4 subfields, corresponding to a survey of
area $A = 102.4^{\prime 2}$, and average over the resulting 144
subfields. Adjacent subfields do not represent fully independent
realizations of the convergence field, but the correlations are
negligible: when averaging over only a bootstrapped subset of
subfields, we get no systematic deviation but only a noisier estimate
of the covariance. Note that because of the splitting, the maximum
usable aperture radius is now 17 arc minutes.

We take into account apertures with centers not closer to the border
than three times the aperture radius $\theta$. This results in an
effective area $A_{\rm eff}(\theta)$ which is smaller than the
original area $A=a^2$, namely $A_{\rm eff}(\theta) = (a - 6
\theta)^2$. Since the covariance is anti-proportional to the observed
area, we can easily apply a correction scheme, and multiply each
covariance matrix entry ${\rm Cov}(\theta_1, \theta_2)$ by
$\sqrt{A_{\rm eff}(\theta_1) A_{\rm eff}(\theta_2)}/A$ in the case of
$\average{M_{\rm ap}^2}$ and $\langle M_{\rm ap, d}^3 \rangle$. For
the generalized third-order aperture mass, where each matrix element
corresponds to two triplets of aperture radii, the effective area
corresponding to the maximum radius of each triplet is inserted into
the correction factor. This correction makes sure that the covariance
matrix corresponds to the same survey area $A$ for all aperture radii.

For the Fisher matrix analysis of constraints on cosmological
parameters (Sect.~\ref{sec:constraints}), we scale the covariances,
obtained from the 2.9 square degree fields, to a corresponding survey
area of 29 square degree, by dividing them by 10, making use of their
$1/A$-dependence.  Note that this increase of survey area is not
equivalent of extending a single patch on the sky, since this
additional observed area will not sample independent but correlated
parts of the large-scale structure and the decrease in cosmic variance
will be less than the increase in area. Our scaling of the area
corresponds to observing 10 independent lines of sight, each one 2.9
square in area.

\subsection{Adding intrinsic ellipticities}

In order to realistically model the noise coming from the intrinsic
ellipticities of the source galaxies, one would have to add a random
ellipticity to each shear value. It has been shown that this is
equivalent to adding a noise term to the convergence $\kappa$
\citep{2000MNRAS.313..524V}. For mass reconstructions, this noise has
to be added to a smoothed $\kappa$ map, however, in our case, no
smoothing is required, thus, to each pixel of $\kappa$ we add a random
Gaussian variable with dispersion $\sigma_\ve = 0.3/\sqrt{2}$. In the
case of $\average{M_{\rm ap}^2}$, this yields the predicted amplitude
to the variance without need of smoothing, as can be seen in
Fig.~\ref{fig:cov2_terms}. The shot-noise contribution to the variance
is in good agreement with the Monte-Carlo method from \cite{KS04}.

The shot-noise term of the variance of $M_{\rm ap}^3$ agrees very well
with the analytical expectation (\ref{varMap3d}), except for large
$\theta$, where only few apertures can be placed onto the field which are
not too close to the border. Apparently, adding intrinsic random
ellipticities to each grid point without smoothing introduces no
artefacts.

\subsection{Gaussianized fields}

For Gaussian random fields, \cite{SvWKM02} found analytic expressions
for the covariance of ${M_{\rm ap}^2}$, which were integrated
via a Monte-Carlo method by \cite{KS04}. In order to compare the
results presented in this work with the Monte-Carlo approach as a
sanity check, we transform the ray-tracing simulations into Gaussian
fields without changing the power spectrum. This is achieved by
multiplying the Fourier transform $\hat \kappa$ of each convergence
field by random phases (destroying the phase correlations). Then for
each Fourier mode $\vec k$, we pick a $\hat \kappa(\vec k)$ value
randomly from one of the 36 fields.

Destroying the phase correlations for each individual field
independently would not have led to the desired goal. Randomizing the
phases cancels the connected 4-point term (kurtosis) of each
individual realization, but not the kurtosis of the underlying
ensemble. Our estimator of the covariance is independent of the
kurtosis of each individual realization, because we first determine
$\langle M_{\rm ap}^2 \rangle$ for each field and then average the
square of this quantity over all fields -- thus for this averaging,
only second-order quantities are taken into account. The process of
remixing the $\hat \kappa$-fields in Fourier space annihilates the
kurtosis of the underlying ensemble, and the resulting fields
represent realizations of a Gaussian random field.

In Fig.~\ref{fig:cov2_terms}, the variance (diagonal of the
covariance) of $ M_{\rm ap}^2$ is plotted. The results from this work
are in fairly good agreement with the \citet{KS04} Monte-Carlo method,
although the cosmic variance term from the ray-tracings is slightly
higher than the one from the Monte-Carlo method.

\begin{figure}[!t]
\begin{center}
\resizebox{0.7\hsize}{!}{
\includegraphics{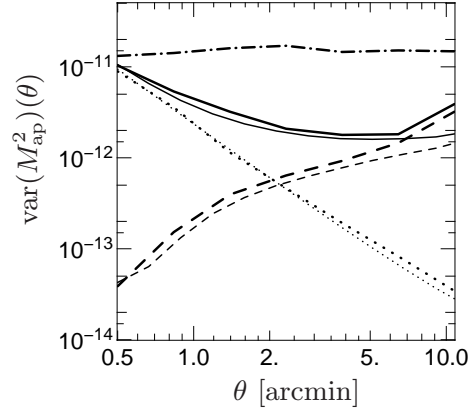}
}
\end{center}
\caption{The variance of $\average{M_{\rm ap}^2}$ from the
Gaussianized $\Lambda$CDM simulations (bold lines), in comparison with
the Monte-Carlo method from \cite{KS04} (thin lines), for a survey
area of $A = {102.4^\prime}^2$. The dotted lines give the variance
from shot noise only (due to the intrinsic ellipticity dispersion), the dashed
curves correspond to cosmic variance only. The solid line include both
error sources; note that it is not the sum of the other two curves --
there is a non-vanishing mixed term. The dash-dotted line
indicates the cosmic variance term for the non-Gaussian case.  }
\label{fig:cov2_terms}
\end{figure}

\begin{figure}[!tb]
\resizebox{\hsize}{!}{
\includegraphics{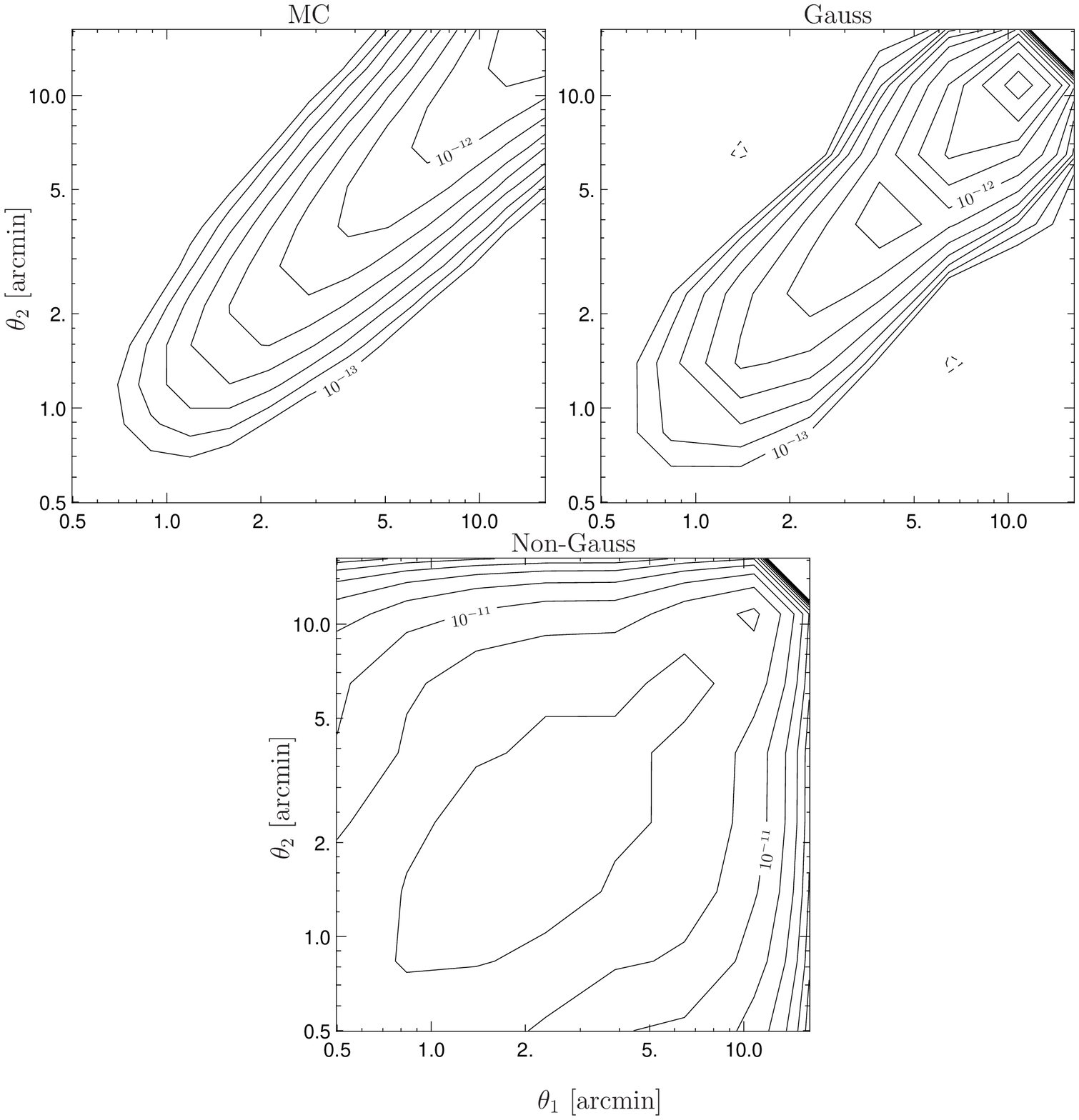}}
\caption{Contour plots of the cosmic-variance-only term of the
covariance of ${M_{\rm ap}^2}$, for the Monte-Carlo method (upper left
panel), the Gaussianized ray-tracing fields (upper right) and the
original fields (lower panel), for a survey area of
$A={102.4^\prime}^2$. The contours are logarithmically spaced.}
\label{fig:cov2-cont}
\end{figure}

It is clear from this figure that non-Gaussianity increases the noise
level on the diagonal by an enormous amount, about two orders of
magnitude at $\sim 1^\prime$. The ratio of the
non-Gaussian to the Gaussian variance is $\propto \theta^{-2}$ for
small $\theta$ and gets less steep for larger $\theta$.

On nearly all scales, cosmic variance dominates
the shot noise. From Fig.~\ref{fig:cov2-cont}, we see that due
to mode-coupling, high cross-correlations between different angular
scales are introduced, present on the off-diagonal of the covariance.

\subsection{The case of $\average{M_{\rm ap}^3}$}
\label{sec:map3}

As for the second-order case, the variance of the aperture mass
skewness, ${\rm var}(M_{\rm ap, d}^3(\theta)) = {\rm var}(M_{\rm
ap}^3(\theta, \theta, \theta))$
is dominated by cosmic variance which is larger than the shot noise on
all but very small scales.

The covariance matrix of ${M_{\rm ap}^3}$ is not diagonal-dominant,
and shows a self-similar pattern with many secondary diagonals,
originating from the reordering of $\langle M_{\rm ap}^3(\theta_1,
\theta_2, \theta_3)\rangle$ into a single vector, which inevitably
creates repeating entries of similar combinations of aperture radii.
The correlation of $\langle M_{\rm ap}^2 \rangle$ for two aperture
radii $\theta_1 \ge \theta_2$ is a quickly decreasing function of the
ratio $\theta_1 / \theta_2$.  In the case of $\langle M_{\rm ap}^3
\rangle$ however, there are many combinations of filter scales which
show a high correlation. This fact together with the small sample of
realizations of $\kappa$-fields causes the covariance matrix to be
very ill-conditioned. For our Fisher matrix analysis
(Sect.~\ref{sec:fisher}), we have to invert the covariance matrix. We
find stable results for the matrix inverting when the ratio of
adjacent aperture radii is chosen not to be too small, i.e.\ larger
than about 1.5.


One way to determine whether our estimate of the covariance of
${M_{\rm ap}^3}$ is reasonable would involve 6-point statistics, which
is not feasible analytically. Instead, we slightly modify the aperture
radii used in the analysis and get a rough estimate of the accuracy of
this method. We comment on the stability of our results in
Sect.~\ref{sec:stability}.

\section{Constraints on cosmological parameters}
\label{sec:constraints}

From the simulated data, we ``observe'' a data vector $\vec M$, which
in our case consists of the values of $\average{M_{\rm ap}^2}$ and/or
$\average{M_{\rm ap}^3}$ as a function of angular scales. Using a
theoretical model, and approximating our observables as Gaussian
variables, we construct a likelihood function ${\cal L}(\vec M; \vec
p)$, which depends on a number of model parameters $\vec p = (p_1,
p_2, \ldots, p_n)$. The likelihood is ${\cal L} \propto \exp(-\chi^2/2)$
with
\begin{eqnarray}
\chi^2(\vec M; \vec p) = \lefteqn{\sum_{kl} \left( M_k(\vec p) -
    M_k(\vec p_0) \right) ({\rm Cov}^{-1})_{kl}} \nonumber \\ &
    & \times \left( M_l(\vec p) - M_l(\vec p_0) \right).
\label{chi}
\end{eqnarray}
where the indices $k$ and $l$ run over the individual data points.

\subsection{The input data}
\label{sec:input}

We distinguish the following five cases for the input data vector $\vec M$
and its covariance $\Cov$:
\begin{enumerate}
\item (`2') $M_l = \average{M_{\rm ap}^2(\theta_l)}$, $\Cov$ =
  covariance of $M_{\rm ap}^2$.

\item (`3') $M_l = \langle{M_{\rm ap, d}^3(\theta_l)}\rangle$, $\Cov$ =
  covariance of $M_{\rm ap,d}^3$.

\item (`3d') $M_l = \average{M_{\rm ap}^3(\theta_i, \theta_j, \theta_k)}$ for
  a combination of three filter radii which after relabeling
  corresponds to index $l$ as described in
  Sect.~\ref{sec:cov-not}. $\Cov$ = covariance of $M_{\rm ap}^3$.
  
\item (`2+3d') $M_l = $ some element from the concatenated data vector
  containing $\average{M_{\rm ap}^2}$ and $\average{M_{\rm ap, d}^3}$.
  $\Cov$ is a block matrix containing the covariances of $M_{\rm
    ap}^2$ and ${M_{\rm ap, d}^3}$ on the diagonal and the
  cross-correlation on the off-diagonal.

\item (`2+3') $M_l = $ some element from the concatenated data vector
  containing $\average{M_{\rm ap}^2}$ and $\average{M_{\rm
  ap}^3}$. $\Cov$ is a block matrix containing the covariances of
  $M_{\rm ap}^2$ and ${M_{\rm ap}^3}$ on the diagonal and the
  cross-correlation on the off-diagonal.
\end{enumerate}

Our choice of the survey geometry corresponds to ten uncorrelated
fields, each of size $A=102.4^{\prime 2}$. In order to get the
covariances of the aperture mass statistics, we split each of the 36
ray-tracing simulations into four subfields and calculate the rms over
the 144 resulting fields. We devide the resulting covariance matrices
by a factor of 10, which then correspond to a total survey area of
\mbox{$A=10 \cdot 102.4^{\prime 2} = 29$} square degree.  We use six
different filter radii, logarithmically spaced between 1 and 15 arc
minutes and thus have six data points for each of $\langle M_{\rm
ap}^2 \rangle$ and $\langle M_{\rm ap, d}^3 \rangle$ and 56 for
$\langle M_{\rm ap}^3 \rangle$.

\subsection{Fisher matrix}
\label{sec:fisher}

It would be desirable to calculate the full
$n$-dimensional likelihood function in order to make predictions about
error bars and directions of degeneracies between parameters.  This,
however, is extremely time-consuming even for sparse sampling in
parameter space, because for every $\vec p$, the bispectrum and the
aperture mass statistics have to be calculated involving
three-dimensional integrals.

Instead, we use the Fisher information matrix \citep{KS69,TTH97} which
gives us a local description of the likelihood ${\cal L}$ at its
maximum.  The Fisher matrix is defined as

\begin{equation}
\Vec{F}_{ij} = \left\langle\frac{\del^2[-\ln \cal L]}{\del p_i \del
    p_j}\right\rangle
 = \left(\frac{\del^2[-\ln \cal L]}{\del p_i \del
    p_j}\right)_{\vec p = \vec p_0},
\label{fisher}
\end{equation}
where $\vec p_0$ denotes the ``true'' parameter values, in our case
the input parameters of the simulations, see Table
\ref{tab:param}. The second equality in (\ref{fisher}) holds if the
maximum likelihood estimator of $\vec p_0$ is unbiased.  The Fisher
matrix is the expectation value of the Hessian matrix of $({-\ln \cal
L})$ at $\vec p = \vec p_0$, which in the case of an unbiased maximum
likelihood estimator coincides on average with $\cal L$'s maximum --
thus it is a measure of how fast ${\cal L}$ falls off from the
maximum.

The smallest possible variance $\sigma$ of any unbiased estimator of
some parameter $p_i$ is given by the Cram\'er-Rao inequality
\begin{equation}
\sigma(p_i) \ge \sqrt{(F^{-1})_{ii}} \; ;
\end{equation}
the expression on the right-hand side is called the \emph{minimum
variance bound} (MVB).

Under the assumption that the parameter dependence of the covariance
can be neglected, we get from eq.\ (\ref{chi}) and (\ref{fisher}):
\begin{equation}
F_{ij} = \sum_{kl} \left(\frac{\del M_k}{\del p_i}\right)
      \left({\rm Cov}^{-1}\right)_{kl} \left(\frac{\del
      M_l}{\del p_j}\right).
\label{fisher-final}
\end{equation}
The derivatives of the aperture mass statistics with respect to the
parameters $p_i$ are calculated numerically from the HEPT model, using
polynomial extrapolation of finite differences \citep{nr}. The Fisher
matrix for the four combinations of second- and third-order statistics
considered in this work is given in Table \ref{tab:fisher}.

\begin{table*}
\label{tab:fisher}
\caption{Fisher matrix for the five different input data as listed in
Sect.\ \ref{sec:input}, denoted by `2', `3', `3d', `2+3d' and `2+3',
respectively. The survey is 29 square degree, all entries are given in
units of $10^4$.}
\begin{center}
\begin{tabular}{c|r|rrrrrr}\hline\hline
& & $\Omegam$ & $\Omega_\Lambda$ & $\Gamma$ & $\sigma_8$ & $n_{\rm s}$
 &  $z_0$ \\ \hline
& $\Omegam$                  & $ 1.766$ & $-0.268$ & $ 1.543$ & $ 1.597$ & $ 0.601$ & $ 1.079$ \\
& $\Omega_\Lambda$           & $-0.268$ & $ 0.239$ & $-0.784$ & $-0.424$ & $-0.336$ & $-0.413$ \\
& $\Gamma$                   & $ 1.543$ & $-0.784$ & $ 2.875$ & $ 1.905$ & $ 1.202$ & $ 1.630$ \\
\raisebox{1.5ex}[-1.5ex]{2}%
& $\sigma_8$                 & $ 1.597$ & $-0.424$ & $ 1.905$ & $ 1.618$ & $ 0.766$ & $ 1.200$ \\
& $n_{\rm s}$                & $ 0.601$ & $-0.336$ & $ 1.202$ & $ 0.766$ & $ 0.507$ & $ 0.674$ \\
& $z_0$                      & $ 1.079$ & $-0.413$ & $ 1.630$ & $ 1.200$ & $ 0.674$ & $ 0.975$ \\
\hline                                                                                            
& $\Omegam$                  & $ 1.698$ & $-0.315$ & $ 1.260$ & $ 1.848$ & $ 0.478$ & $ 0.661$ \\ 
& $\Omega_\Lambda$           & $-0.315$ & $ 0.445$ & $-1.304$ & $-0.852$ & $-0.552$ & $-0.452$ \\ 
& $\Gamma$                   & $ 1.260$ & $-1.304$ & $ 3.952$ & $ 2.803$ & $ 1.653$ & $ 1.389$ \\
\raisebox{1.5ex}[-1.5ex]{3}%
& $\sigma_8$                 & $ 1.848$ & $-0.852$ & $ 2.803$ & $ 2.705$ & $ 1.137$ & $ 1.146$ \\ 
& $n_{\rm s}$                & $ 0.478$ & $-0.552$ & $ 1.653$ & $ 1.137$ & $ 0.701$ & $ 0.579$ \\ 
& $z_0$                      & $ 0.661$ & $-0.452$ & $ 1.389$ & $ 1.146$ & $ 0.579$ & $ 0.544$ \\
\hline                                                                                            
& $\Omegam$                  & $ 0.263$ & $-0.079$ & $ 0.291$ & $ 0.342$ & $ 0.109$ & $ 0.119$ \\ 
& $\Omega_\Lambda$           & $-0.079$ & $ 0.146$ & $-0.422$ & $-0.255$ & $-0.177$ & $-0.141$ \\ 
& $\Gamma$                   & $ 0.291$ & $-0.422$ & $ 1.236$ & $ 0.794$ & $ 0.513$ & $ 0.418$ \\ 
\raisebox{1.5ex}[-1.5ex]{3d}%
& $\sigma_8$                 & $ 0.342$ & $-0.255$ & $ 0.794$ & $ 0.635$ & $ 0.319$ & $ 0.284$ \\ 
& $n_{\rm s}$                & $ 0.109$ & $-0.177$ & $ 0.513$ & $ 0.319$ & $ 0.214$ & $ 0.173$ \\ 
& $z_0$                      & $ 0.119$ & $-0.141$ & $ 0.418$ & $ 0.284$ & $ 0.173$ & $ 0.144$ \\ 
\hline                                                                                            
& $\Omegam$                  & $ 3.256$ & $-0.071$ & $ 1.844$ & $ 2.322$ & $ 0.713$ & $ 1.757$ \\
& $\Omega_\Lambda$           & $-0.071$ & $ 0.363$ & $-0.999$ & $-0.466$ & $-0.434$ & $-0.379$ \\
& $\Gamma$                   & $ 1.844$ & $-0.999$ & $ 3.613$ & $ 2.405$ & $ 1.527$ & $ 1.924$ \\
\raisebox{1.5ex}[-1.5ex]{2+3d}%
& $\sigma_8$                 & $ 2.322$ & $-0.466$ & $ 2.405$ & $ 2.180$ & $ 0.974$ & $ 1.592$ \\
& $n_{\rm s}$                & $ 0.713$ & $-0.434$ & $ 1.527$ & $ 0.974$ & $ 0.653$ & $ 0.799$ \\
& $z_0$                      & $ 1.757$ & $-0.379$ & $ 1.924$ & $ 1.592$ & $ 0.799$ & $ 1.337$ \\
\hline                                                                                            
& $\Omegam$                  & $ 7.046$ & $ 0.348$ & $ 2.457$ & $ 4.440$ & $ 0.939$ & $ 3.219$ \\ 
& $\Omega_\Lambda$           & $ 0.348$ & $ 0.764$ & $-1.775$ & $-0.740$ & $-0.773$ & $-0.435$ \\ 
& $\Gamma$                   & $ 2.457$ & $-1.775$ & $ 5.846$ & $ 3.925$ & $ 2.486$ & $ 2.756$ \\ 
\raisebox{1.5ex}[-1.5ex]{2+3}%
& $\sigma_8$                 & $ 4.440$ & $-0.740$ & $ 3.925$ & $ 4.162$ & $ 1.597$ & $ 2.635$ \\ 
& $n_{\rm s}$                & $ 0.939$ & $-0.773$ & $ 2.486$ & $ 1.597$ & $ 1.076$ & $ 1.159$ \\ 
& $z_0$                      & $ 3.219$ & $-0.435$ & $ 2.756$ & $ 2.635$ & $ 1.159$ & $ 2.169$ \\
\hline\hline
\end{tabular}
\end{center}
\end{table*}

\subsection{Minimum Variance Bounds (MVBs)}

For various combinations of cosmological parameters, we compute the
MVBs from the Fisher information matrix (\ref{fisher-final}). As the
covariance scales with $A^{-1}$ (where $A$ is the observed area), the
MVB is roughly proportional to $1/\sqrt{A}$. First, the analysis is
done for only two parameters, in order to graphically display the
MVBs. Then,
simultaneous MVBs for three and more parameters are calculated.

\subsubsection{Two parameters}

In Fig.~\ref{fig:2dellipses}, we show the MVBs as ellipses in
two-dimensional subspaces of the parameter space. The hidden
parameters are fixed. In all cases, the combination of
$\average{M_{\rm ap}^2}$ and $\average{M_{\rm ap}^3}$ leads to a
substantial reduction in the 1-$\sigma$-error. As expected, the
generalized third-order aperture mass statistics yields much better
constraints than the `diagonal' version $\langle{M_{\rm ap,
d}^3}\rangle$. The direction of degeneracy is slightly different for
some parameter pairs, most notably when the source redshift parameter
$z_0$ is involved, making the combination of the statistics very
effective in these cases. The $\Omegam$-$\sigma_8$-degeneracy is
lifted partially and the combined Fisher matrix analysis yields a
large improvement on the error of the two parameters. Contrary to
that, the pair $(\Gamma, n_{\rm s})$ is degenerate to a high level for
both $\average{M_{\rm ap}^2}$ and $\average{M_{\rm ap}^3}$ as well as
for their combination.

Note that the combined 1-$\sigma$-errors are not completely determined
by the product of the likelihoods of $\average{M_{\rm ap}^2}$ and
$\average{M_{\rm ap}^3}$. The combined covariance is not the direct
product of the covariances of $\average{M_{\rm ap}^2}$ and
$\average{M_{\rm ap}^3}$ because of the contribution from the
cross-correlation between both statistics.

It is not surprising that the directions of degeneracy between
parameters are more or less similar for $\average{M_{\rm ap}^2}$ and
$\average{M_{\rm ap}^3}$, with larger differences existing when $z_0$
is one of the free parameters. Both statistics depend on the convergence
power spectrum, because in HEPT as well as in quasi-linear PT, the
bispectrum of the matter fluctuations is given in terms of the power
spectrum (\ref{bkappabar}). The differences between $\average{M_{\rm
ap}^2}$ and $\average{M_{\rm ap}^3}$ mainly come from their different
dependence on the projection prefactor and the lens efficiency $G$
(eq.~\ref{G}). The projection is most sensitive to the source
redshift, and of all parameters, changes in $z_0$ show up in a most
distinct way for $\average{M_{\rm ap}^2}$ and $\average{M_{\rm
ap}^3}$.

Since the degeneracy directions between $\average{M_{\rm ap}^2}$ and
the skewness $\average{M_{\rm ap, d}^3}$ are very similar, not much
improvement is obtained when these two statistics are combined and
therefore, the corresponding error ellipses are not drawn in
Fig.~\ref{fig:2dellipses}.

\begin{figure*}[ht!]
\resizebox{\hsize}{!}{
\includegraphics{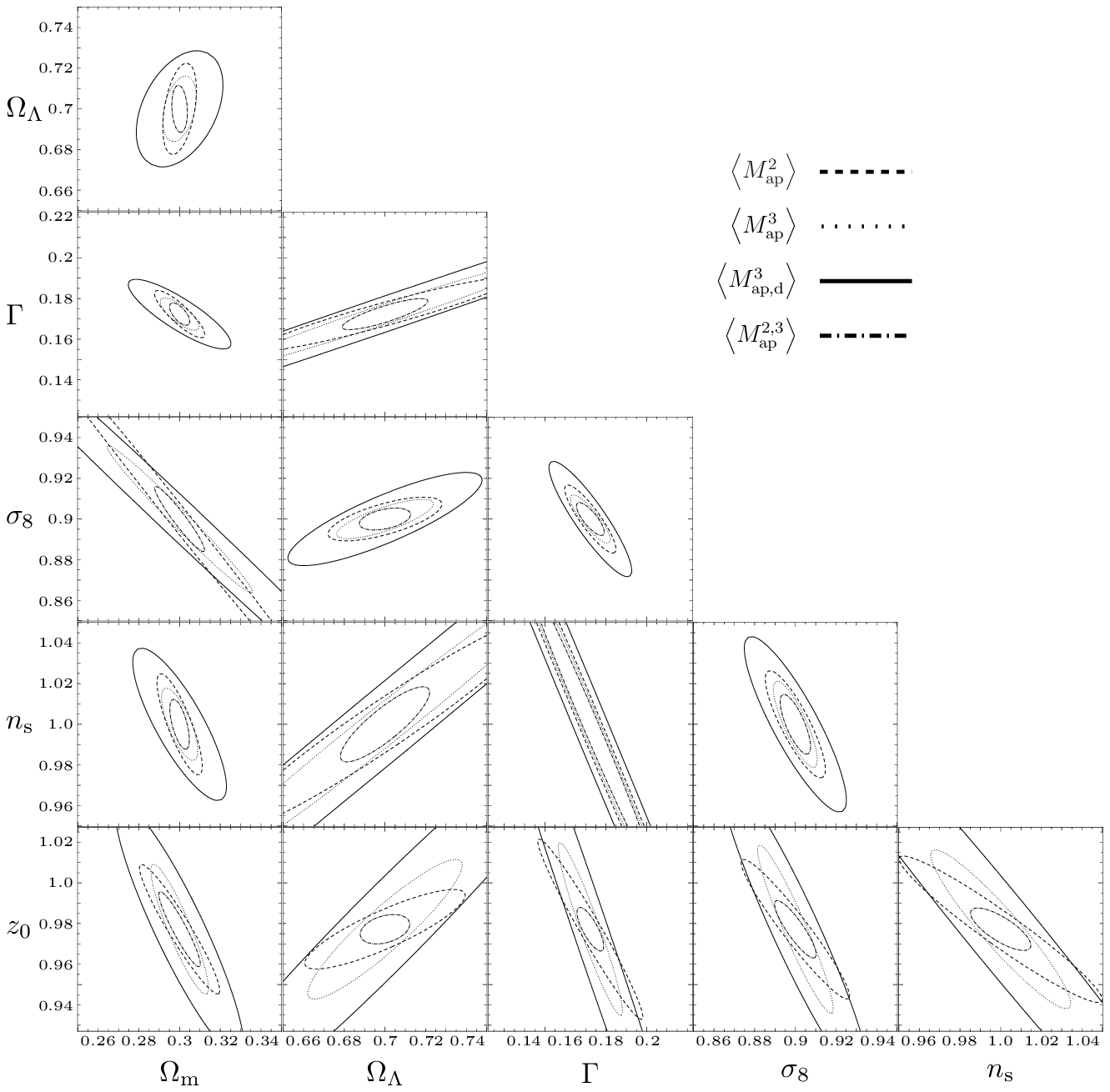}}
\caption{1-$\sigma$-error ellipses from the Fisher matrix. The hidden
  parameters are kept fixed. Dashed line: $\average{M_{\rm ap}^2}$,
  dotted line: $\average{M_{\rm ap}^3}$, solid line: $\langle{M_{\rm
  ap,d}^3}\rangle$, dash-dotted line: combination of $\average{M_{\rm
  ap}^2}$ and $\average{M_{\rm ap}^3}$ as described in
  Sect.~\ref{sec:input}. If one of the parameters is $\Omegam$, a flat
  Universe is assumed (except for the
  $\Omegam$-$\Omega_\Lambda$-plot). The assumed survey area is
  29 square degree.}
\label{fig:2dellipses}
\end{figure*}

\subsubsection{Three and more parameters}
\label{sec:more-d}

We calculate the MVBs for three and more parameters simultaneously for
various combinations of parameters and for each input data as
described in Sect.~\ref{sec:input}.
The results are given in Table \ref{tab:mvb}.  All hidden parameters
are fixed to their fiducial values, see Table \ref{tab:param}. If not
both $\Omegam$ and $\Omega_\Lambda$ vary, a flat Universe is assumed.

In most of the cases, the error bars from the generalized third-order
aperture mass statistics $\langle M_{\rm ap}^3 \rangle$ are smaller
than those from its second-order counterpart $\langle M_{\rm ap}^2
\rangle$. This trend gets stronger the more free cosmological
parameters are involved, since the measurement of $\langle M_{\rm
ap}^3 \rangle$ provides more data points and therefore more degrees of
freedom\footnote{This is true only to some extent since the data
points are correlated.}. The skewness of the aperture mass $\langle
M_{\rm ap, d}^3 \rangle$ yields by far the worst constraints on the
parameters.

In all of the cases, the combination of $\langle M_{\rm ap}^2 \rangle$
and $\langle M_{\rm ap}^3 \rangle$ results in an improvement on the
parameter constraints. This improvement can be rather small, e.g.\ in
the cases when both $\Gamma$ and $n_{\rm s}$ are involved. Then the
combined MVB is dominated mainly by the MVB of $\langle M_{\rm ap}^3
\rangle$, and the additional information from $\langle M_{\rm ap}^2
\rangle$ is unimportant. However, for a number of parameter
combinations, the combined error represents an improvement of a factor
two and more, indicating that the dependence of the two statistics on
the cosmological parameters is different to some degree, and their
combination lifts the degeneracy substantially. Amongst other, this
occurs for the pair $\Omegam$ and $\sigma_8$. Even if a rather good
constraint on these two parameters from $\langle M_{\rm ap}^3 \rangle$
is combined with a large MVB, the combined error can be reduced by a
factor of two and more, thus the most prominent parameter degeneracy
for second-order cosmic shear between $\Omegam$ and $\sigma_8$ can
partially be broken by adding third-order statistics.

When $\langle \Mapsq \rangle$ is combined with the generalized
aperture-mass statistics (the case `2+3') and the skewness (`2+3d'),
the first combination always yields better parameter constraints than
the latter. For three free parameters, the first combination is
typically a factor of two better, if more parameters are involved, the
improvement factor is even larger, up to a factor of ten when all six
parameters are free. Thus, the preference of $\langle M_{\rm ap}^3
\rangle$ over the skewness of $\Map$ is justified also when it is
combined with the second-order aperture mass statistics.

In general, constraints on the cosmological constant $\Omega_\Lambda$
are weaker than for the other parameters, and although the combination
of second- and third order aperture mass statistics gives some
improvement on the error, $\Omega_\Lambda$ remains the least known
parameter.


\begin{table}
\caption{MVBs for various combinations of three and more cosmological
parameters, corresponding to a 29 square degree survey. The hidden
parameters are kept fixed. `2', `3', `3d', `2+3d' and `2+3'
stand for the five different input data as described in Sect.\
\ref{sec:input}. If $\Omega_\Lambda$ is not a free parameter,
a flat Universe is assumed.}
\begin{center}
\begin{tabular}{c|cccccc}\hline\hline
 & $\Omegam$ & $\Omega_\Lambda$ & $\Gamma$ & $\sigma_8$ & $n_{\rm s}$
 &  $z_0$ \\ \hline
2   & 0.077 &       &       & 0.104 &       & 0.035 \\
3   & 0.041 &       &       & 0.053 &       & 0.047 \\
3d  & 0.189 &       &       & 0.219 &       & 0.116 \\
2+3d& 0.027 &       &       & 0.030 &       & 0.029 \\
2+3 & 0.016 &       &       & 0.016 &       & 0.019 \\ \hline
2   & 0.087 &       & 0.015 & 0.119 &       &      \\
3   & 0.063 &       & 0.017 & 0.078 &       &      \\
3d  & 0.267 &       & 0.043 & 0.300 &       &      \\
2+3d& 0.027 &       & 0.012 & 0.044 &       &      \\
2+3 & 0.015 &       & 0.008 & 0.024 &       &      \\ \hline
2   & 0.083 &       &       & 0.110 & 0.029 &     \\
3   & 0.059 &       &       & 0.072 & 0.035 &     \\
3d  & 0.222 &       &       & 0.242 & 0.077 &     \\
2+3d& 0.026 &       &       & 0.040 & 0.025 &      \\
2+3 & 0.014 &       &       & 0.022 & 0.017 &     \\ \hline
2   & 0.017 &       & 0.093 &       & 0.200 &      \\
3   & 0.010 &       & 0.051 &       & 0.113 &      \\
3d  & 0.343 &       & 0.158 &       & 0.347 &      \\
2+3d& 0.010 &       & 0.063 &       & 0.138 &      \\
2+3 & 0.005 &       & 0.035 &       & 0.078 &      \\ \hline
2   & 0.087 & 0.106 &       & 0.113 &       &      \\
3   & 0.057 & 0.089 &       & 0.067 &       &      \\
3d  & 0.244 & 0.328 &       & 0.263 &       &      \\
2+3d& 0.033 & 0.056 &       & 0.047 &       &      \\
2+3 & 0.019 & 0.037 &       & 0.028 &       &      \\ \hline
2   & 0.095 & 0.744 & 0.353 & 0.285 &       &      \\
3   & 0.065 & 0.117 & 0.053 & 0.085 &       &      \\
3d  & 0.387 & 0.817 & 0.490 & 0.542 &       &      \\
2+3d& 0.066 & 0.243 & 0.078 & 0.053 &       &      \\
2+3 & 0.028 & 0.088 & 0.026 & 0.030 &       &      \\ \hline
2   & 0.569 &       & 0.270 & 0.552 &       & 0.645 \\
3   & 0.080 &       & 0.033 & 0.088 &       & 0.091 \\
3d  & 1.113 &       & 0.498 & 1.084 &       & 1.332 \\
2+3d& 0.218 &       & 0.090 & 0.224 &       & 0.213 \\
2+3 & 0.069 &       & 0.031 & 0.072 &       & 0.069 \\ \hline
2   & 0.674 & 2.447 & 2.199 & 1.857 & 1.618 &      \\
3   & 0.065 & 0.127 & 0.065 & 0.085 & 0.125 &      \\
3d  & 7.306 & 1.349 & 7.613 & 9.686 & 8.344 &      \\
2+3d& 0.075 & 0.263 & 0.101 & 0.070 & 0.186 &      \\
2+3 & 0.031 & 0.094 & 0.043 & 0.034 & 0.090 &      \\ \hline
2   & 3.554 & 5.535 & 2.677 & 3.719 & 2.169 & 3.553 \\
3   & 0.110 & 0.328 & 0.084 & 0.092 & 0.140 & 0.220 \\
3d  & 7.532 & 23.48 & 11.87 & 10.43 & 13.15 & 17.83 \\
2+3d& 0.928 & 1.101 & 0.620 & 0.996 & 0.701 & 0.812 \\
2+3 & 0.079 & 0.124 & 0.050 & 0.080 & 0.090 & 0.072 \\
\hline\hline
\end{tabular}
\end{center}
\label{tab:mvb}
\end{table}


\subsection{Correlation between parameters}
\label{sec:corr}

The correlation coefficient of the inverse Fisher matrix
\begin{equation}
r_{ij} = \frac{F^{-1}_{ij}}{\sqrt{ F^{-1}_{ii} F^{-1}_{jj} }}
\label{corr}
\end{equation}
is a measure of the correlation between the $i^{\rm th}$ and $j^{\rm
th}$ parameter. For $i \ne j$, it can vary between -1 and 1. In the
two-dimensional case, $r_{12}=r_{21}=0$ corresponds to an error ellipse
with major and minor axes parallel to the coordinate (parameter) axes
-- the probability distribution of the parameters factorizes.  For
$r_{12} \rightarrow 1$, the ellipse degenerates to a line.

Table \ref{tab:corr} shows the correlation coefficient between all
cosmological parameters considered in this work. For the combination
of $\langle \Mapsq \rangle$ and $\langle M_{\rm ap, d}^3 \rangle$
(`2+3d'), the correlation is very large for all parameter pairs, the
difference to unity in some cases is only of the order of $10^{-3}$.
The degeneracy directions of $\langle \Mapsq \rangle$ and $\langle
M_{\rm ap, d}^3 \rangle$ are very similar, thus the combination of the
two causes the correlation between parameters to be very high.

\begin{table}[!tb]
\caption{The correlation coefficient $r_{ij}$ (\ref{corr}) of the
  inverse Fisher matrix (\ref{corr}).  `2', `3', `3d', `2+3d' and
  `2+3' stand for the five different input data as described in
  Sect.\ \ref{sec:input}. Note that the correlation matrix $r$ is
  symmetric and unity on the diagonal.}
\begin{center}
\begin{tabular}{c|c|rrrrr}\hline\hline
& & $\Omega_\Lambda$ & $\Gamma$ & $\sigma_8$ & $n_{\rm s}$ & $z_0$ \\ \hline
& $\Omegam$        & $-$0.80 &    0.71 & $-$0.94 & $-$0.79 & $-$0.98 \\
& $\Omega_\Lambda$ &     & $-$0.15 &    0.56 &    0.28 &    0.90 \\
& $\Gamma$         &         &     & $-$0.90 & $-$0.98 & $-$0.57 \\
\raisebox{1.5ex}[-1.5ex]{2}%
& $\sigma_8$       &         &         &     &    0.95 &    0.87 \\
& $n_{\rm s}$      &         &         &         &     &    0.67 \\ \hline
& $\Omegam$        &   $-$0.81 &   $-$0.31 &   $-$0.84 &    0.35 &   $-$0.81 \\
& $\Omega_\Lambda$ &     &    0.66 &    0.41 &   $-$0.29 &    0.92 \\
& $\Gamma$         &         &     &   $-$0.15 &   $-$0.69 &    0.63 \\
\raisebox{1.5ex}[-1.5ex]{3}%
& $\sigma_8$       &         &         &     &   $-$0.14 &    0.38 \\
& $n_{\rm s}$      &         &         &         &     &   $-$0.46 \\ \hline
& $\Omegam$        &    $-$0.29 &    0.43 &   $-$0.81 &   $-$0.43 &   $-$0.24 \\
& $\Omega_\Lambda$ &    &    0.74 &   $-$0.33 &   $-$0.74 &    1.00 \\
& $\Gamma$         &         &    &   $-$0.88 &   $-$1.00 &    0.77 \\
\raisebox{1.5ex}[-1.5ex]{3d}%
& $\sigma_8$       &          &         &     &    0.87 &   $-$0.37 \\
& $n_{\rm s}$      &           &         &         &     &   $-$0.77 \\ \hline
& $\Omegam$        &  $-$0.99 &    0.98   &   $-$1.00 &   $-$0.97 &   $-$1.00 \\
& $\Omega_\Lambda$ &          &   $-$0.94 &    0.98 &    0.96 &    0.97 \\
& $\Gamma$         &          &           & $-$0.99 &   $-$0.98 &   $-$0.99 \\
\raisebox{1.5ex}[-1.5ex]{2+3d}%
& $\sigma_8$       &          &           &         &    0.97 &    1.00 \\
& $n_{\rm s}$      &          &           &         &         &   0.96  \\ \hline
& $\Omegam$        &    $-$0.87 &    0.46 &   $-$0.99 &   $-$0.12 &   $-$0.92 \\
& $\Omega_\Lambda$ &     &   $-$0.20 &    0.83 &    0.24 &    0.65 \\
& $\Gamma$         &         &     &   $-$0.55 &   $-$0.66 &   $-$0.52 \\
\raisebox{1.5ex}[-1.5ex]{2+3}%
& $\sigma_8$       &       &         &    &    0.17 &    0.90 \\
& $n_{\rm s}$      &       &         &         &    &   $-$0.04 \\ \hline\hline
\end{tabular}
\end{center}
\label{tab:corr}
\end{table}

\subsection{Stability}
\label{sec:stability}

In order to check our Fisher matrix analysis for consistency and
stability towards small changes of the input data, we redo our
calculations with slightly different aperture radii. For changes of a
couple of percent in the aperture radii, the resulting Fisher matrix
elements vary of the order of up to 10 percent. The MVBs (see
Sect.~\ref{sec:fisher}) fluctuate by about the same amount if two or
three parameters are considered to be determined from the data
simultaneously.  However, for four and five free parameters, the MVBs
are less stable, since the Fisher matrix is numerically very
ill-conditioned and the inversion is a non-linear operation. In
general, the MVBs for $\langle M_{\rm ap}^3 \rangle$ are less stable
than the ones for $\langle M_{\rm ap}^2 \rangle$.


The eigenvectors of $F_{ij}^{-1}$ are less affected by a different
sampling of aperture radius. Angles between original and
modified eigenvectors are typically only a few degree.
The variation of the correlation coefficient $r_{ij}$ (\ref{corr})
is less than $\sim 0.1$ if up to four parameters are considered. For a
higher-dimensional Fisher matrix however, the variation can be higher,
similar to the case of the MVB.

\section{Summary and Conclusion}

The power spectrum of large-scale (dark-)matter fluctuations was until
recently the most important quantity that has been measured --
directly or indirectly -- by cosmic shear. Interesting constraints on
cosmological parameters like $\Omegam$ and $\sigma_8$ have been
obtained from second-order cosmic shear statistics.

The bispectrum of density fluctuations contains complementary
information about structure evolution and cosmology. It is a measure
of the non-Gaussianity of the large-scale structure. Current cosmic
shear surveys are at the detection limit of measuring a non-Gaussian
signal significantly, and future observations will certainly determine
the bispectrum with high accuracy.

Combined measurements of the power and the bispectrum yield additional
constraints on cosmological parameters and partially lift degeneracies
between them. The second- and generalized third-order aperture mass
statistics are local measures of the power and bispectrum,
respectively. In this work, we made predictions about cosmological
parameter estimations from combined measurements of these two weak
lensing statistics. Using $\Lambda$CDM ray-tracing simulations, we
calculated the covariance of $ M_{\rm ap}^2$ and $M_{\rm ap}^3$ and
their cross-correlation. We performed an extensive Fisher matrix
analysis and obtained minimum variance bounds (MBVs) for a variety of
combinations of cosmological parameters.

The generalized third-order aperture mass statistics \citep{SKL04} is
the correlator of $\Map$ for three different aperture radii. In
contrast to the skewness of $\Map$ which probes the bispectrum for
equilateral triangles only, the generalized third-order aperture mass
is in principle sensitive to the bispectrum on the complete
$\ell$-space. Therefore, it contains much more information about
cosmology than the skewness alone.

The direction of degeneracy between the cosmological parameters
considered here are similar for second- and third-order statistics.
However, in most cases the combination of $\langle M_{\rm
ap}^2\rangle$ and $\langle M_{\rm ap}^3\rangle$ gives substantial
improvement on the predicted parameter constraints.
The MVBs decrease by a factor of two or more for most of the
parameter combinations. When the source redshift $z_0$ is not fixed
but also to be determined from the data, the errors on the other
parameters increase and the improvement by combining $\langle
M_{\rm ap}^2\rangle$ and $\langle M_{\rm ap}^3\rangle$ is
lowered.

We combined the second-order aperture mass statistics $\langle \Mapsq
\rangle$ with both the skewness and the generalized third-order
aperture mass. The latter combination gives much better parameter
constraints than the first one. For six parameters to be determined
from the data simultaneously, the corresponding MVBs are better by a
factor of about 10 for each parameter.

The $\Omegam$-$\sigma_8$-degeneracy is very prominent for both the
second- and the third-order statistics of $\Map$
individually. However, by combining the two, the degeneracy is
partially lifted -- the 1-$\sigma$-errors of both parameters drop by a
factor of two or more, depending on which other parameters are also
considered to be determined from the data. The $n_{\rm
s}$-$\Gamma$-degeneracy, however, can not be broken by combining
$\langle M_{\rm ap}^2 \rangle$ and $\langle M_{\rm ap}^3 \rangle$, the
determination of this pair of parameters is dominated by $\langle
M_{\rm ap}^3 \rangle$.

For the given range of 1 to 15 arc minutes for the aperture radii
considered in this work the generalized third-order aperture mass
statistics is dominant for the determination of most of the
cosmological parameter combinations. The measurement error from
$\langle M_{\rm ap}^2 \rangle$ is in general larger than the one from
$\langle M_{\rm ap}^3 \rangle$. However, in most of the cases,
even a weak constraint from $\langle M_{\rm ap}^2 \rangle$ alone
contributes valuable information to the combination of the two
statistics, and the combined error is much smaller than the one from
the individual measurements.

If the range of apertures is extended, would we expect the resulting
improvement on the parameter estimation from the third-order aperture
mass statistics to be higher than from second-order? For the former,
the number of data points increases with the third power of the number
of aperture radii, whereas for the latter, the increase is only
linear. Thus, for an increase in the number of measured apertures, the
constraints using $\langle M_{\rm ap}^3 \rangle$ should improve more
than those from $\langle M_{\rm ap}^2 \rangle$.  On the other hand,
the data points are not at all uncorrelated; in fact, as it is shown
in this work (Sect.~\ref{sec:map3}), the correlation can be very
strong for various combinations of aperture radius triples. Moreover,
for large scales ($\theta \gsim 30^\prime$, see
Fig.~\ref{fig:map2_comp}), the linear regime of the large-scale
structure is probed, where non-Gaussian contributions are small, and
the information content of third-order shear statistics is
diminished. We conclude that angular scales up to about 30 arc minutes
will be a good choice for the measurement of the generalized
third-order aperture mass statistics of cosmic shear. The combination
of this statistics with $\langle M_{\rm ap}^2 \rangle$ will improve
the resulting constraints on cosmological parameters quite substantially.

\begin{acknowledgements}

We thank Takashi Hamana for kindly providing his ray-tracing
simulations, Mike Jarvis for his tree-code algorithm which we used to
calculate the 3PCF and Masahiro Takada, Mike Jarvis, Patrick Simon and
Marco Lombardi for helpful discussions. We are very grateful to the
anonymous referee whose suggestions helped to improve the paper. This
work was supported by the German Ministry for Science and Education
(BMBF) through the DLR under the project 50 OR 0106, and by the
Deutsche Forschungsgemeinschaft under the project SCHN 342/3--1.

\end{acknowledgements}

\begin{appendix}

\section{Covariance of $M_{\rm ap}^3$ for
         shot-noise only}
\label{covmap3sn}

Analogous to \cite{SvWKM02}, we analytically calculate the variance of
$M_{\rm ap}^3$ in the case of shot-noise only, by integrating over the
covariance of the shear 3PCF.  An unbiased estimator of the natural
component $\Gamma^{(0)}$ of the 3PCF \citep{tpcf1} is
\begin{equation}
  \hat \Gamma^{(0)}(T_x) = \frac{1}{N_{\rm T}(T_x)} \sum_{ijk} \enat(ijk)
  \Delta_{T_x}(ijk),
  \label{hatgamma}
\end{equation}
where $T_x$ represents a triangle of points $\vec \vt_i, \vec \vt_j$
and $\vec \vt_k$, uniquely given e.g.\ by two side lengths $x_1, x_2$
and the angle $\vp$ between them. $N_{\rm T}(T_x)$ is the number of
triangles within the bin containing $T_x$, and $\enat(ijk)$ an
estimator of $\Gamma^{(0)}$; it is the following linear combination of
products of ellipticities of three galaxies at positions $\vec \vt_i,
\vec \vt_j$ and $\vec \vt_k$, i.e.\
\begin{eqnarray}
\enat & = &  (\ve_{\rm ttt} - \ve_{{\rm t}\times\times} - \ve_{\times {\rm
    t} \times} - \ve_{\times \times {\rm t}}) + \nonumber \\
& & \ii (\ve_{{\rm t
    t}\times} + \ve_{{\rm t} \times {\rm t}} + \ve_{\times {\rm t t}}
    - \ve_{\times \times \times}),
\label{nat}
\end{eqnarray}
where $\ve_{\mu \nu \lambda} = \ve_{\mu \nu \lambda}(\vec \vt_i, \vec
\vt_j, \vec \vt_k) = \ve_\mu(\vec \vt_i) \ve_\nu(\vec \vt_j)
\ve_\lambda(\vec \vt_j)$ for
$\mu, \nu, \lambda \in  \{ $`${\rm t}$'$,$`$\times$'$ \}$,
see eqs.\ (2) and (19) in \cite{tpcf1}. The summation in
(\ref{hatgamma}) is performed over all possible triples of points
$(\vec \vt_i, \vec \vt_j, \vec \vt_k)$, $\Delta_{T_x}(ijk)$ is unity
if the triangle given by $(\vec \vt_i, \vec \vt_j, \vec \vt_k)$ is in
the same bin as $T_x$, and zero otherwise.

The covariance of $\hat \Gamma^{(0)}$ consists of four terms, which
are proportional to $\sigma_\ve^6$, $\sigma_\ve^4$, $\sigma_\ve^2$ and
$\sigma_\ve^0$, respectively. In the case of vanishing cosmic variance,
only the first term contributes; it reads
\begin{eqnarray}
\lefteqn{{\rm Cov}(\hat \Gamma^{(0)}, \hat \Gamma^{(0)}; T_x, T_y) =
  \frac{1}{N_{\rm T}(T_x) N_{\rm T}(T_y)}} \nonumber \\
   & & \times \sum_{ijklmn} \left\langle
  \enat^{(\rm s)}(ijk) \enat^{(\rm s) *}(lmn) \right\rangle \Delta_{T_x}(ijk)
  \Delta_{T_y}(lmn),
\end{eqnarray}
where the superscript `s' indicates the intrinsic (`source')
ellipticity.

The term in angular brackets is non-zero only if the two triangles
given by $(\vec \vt_i, \vec \vt_j, \vec \vt_k)$ and $(\vec \vt_l, \vec
\vt_m, \vec \vt_n)$ are identical (under the assumption that different
galaxies are intrinsically uncorrelated), and factorizes into a sum of
products of three two-point terms, each of the form $\langle
\ve_\mu(\vec \vt_i) \ve_{\mu^\prime}(\vec \vt_l) \rangle \delta_{il} \delta_{\mu
\mu^\prime}$.
With $\langle \ve_{\rm t} \ve_{\rm t} \rangle = \langle \ve_\times
\ve_\times \rangle = \sigma_\ve^2/2$ and $\langle \ve_{\rm t}
\ve_\times \rangle = 0$, the term in angular brackets
becomes $8 \cdot [\sigma_\ve^2/2]^3 = \sigma_\ve^6$. The sum reduces
to a triple sum over $\Delta_{T_x}(ijk)$ which is just the number of
triangles in the respective bin. Finally, we get
\begin{equation}
{\rm Cov}(\hat \Gamma^{(0)}, \hat \Gamma^{(0)}; T_x, T_y) =
\frac{\sigma_\ve^6}{N_{\rm T}(T_x)} \bar \delta(T_x, T_y),
\end{equation}
where $\bar\delta(T_x, T_y)$ is zero if the two triangles $T_x$ and
$T_y$ are in different bins, and unity otherwise.

The covariance of $M_{\rm ap}^3$ is obtained by integrating over the
covariance of the 3PCF. This can be done analytically for the case
when all six aperture radii are equal (this corresponds to the
variance of $M_{\rm ap, d}^3$) and in the absence of a B-mode. We
write eq.\ ({62}) of \cite{SKL04} in the following way, abbreviating
the integral kernel with $R_\theta$,
\begin{eqnarray}
  \lefteqn{\left\langle M^3_{\rm ap}(\theta_1, \theta_2, \theta_3) \right\rangle =
    \int \dd y_1 \int
    \dd y_2 \int \dd \psi} \nonumber \\
  & & \times \tilde \Gamma^{(0)}_{\rm cen}(y_1, y_2, \psi)
  R_{\theta}(y_1, y_2, \psi).
\end{eqnarray}
where $\tilde \Gamma^{(0)}_{\rm cen}$ denotes the 3PCF in the
center-of-mass projection, expressed as a function of triangle sides
as described in \cite{SKL04}.

Before we proceed, we note that for any function $f$,
\begin{eqnarray}
  \lefteqn{\int \dd y_1 \int
    \dd y_2 \int \dd \psi f(y_1, y_2, \psi) \bar \delta(T_x, T_y)}
  \nonumber \\
  & &  = f(x_1,
  x_2, \vp) \Delta x_1 \Delta x_2 \Delta \vp,
\end{eqnarray}
where $\Delta x_1$, $\Delta x_2$, $\Delta \vp$ are given by the bin
size in which the triangle $T_x$ is situated.

For simplicity, we assume that boundary effects due to the finite
field size can be neglected. Then the number of triangles within the
bin characterized by $(x_1, x_2, \vp)$ is $N_{\rm T}(T_x) = N \cdot
(2\pi x_2 \Delta x_2 \, n) \cdot (\Delta x_1 \Delta \vp \, n) = 2\pi A
n^3 x_1 \Delta x_1 x_2 \Delta x_2 \Delta \vp$, where $N$ is the total
number of galaxies and $n$ is the galaxy density ($n=N/A$ for $A$
being the survey area).
%
Thus,
\begin{equation}
  {\rm var}(M_{\rm ap,d}^3; \theta) =  \frac{\sigma_\ve^6}{2\pi A n^3} 
    \int \frac{\dd x_1}{x_1} \int \frac{\dd x_2}{x_2} \int \dd
  \vp \, R^2_{\theta}(x_1, x_2, \vp).
\end{equation}
Solving the integral, we get
\begin{eqnarray}
\lefteqn{{\rm var}(M_{\rm ap, d}^3; \theta)
= \frac{11}{15552 \, \pi^2} \frac{ \sigma_\ve^6}{ A n^3 \, \theta^4}
   =  10^{-16} \left(\frac{\sigma_\ve}{0.3}\right)^6} \nonumber \\ 
& & \times \left(\frac A {9 \,
  {\rm deg}^2}\right)^{-1} \left(\frac n {25 \, {\rm
    arcmin}^{-2}}\right)^{-3} \left(\frac \theta {\rm
  1 arcmin}\right)^{-4}.
\label{varMap3d}
\end{eqnarray}
Note that the variance of $M_{\rm ap}^2$ \citep{SvWKM02} has the same dependence on
the observed area $A$, but is only quadratically invsere as a function
of both $n$ and $\theta$.

\end{appendix}

\bibliographystyle{aa} \bibliography{astro}

\end{document}